\documentclass[useAMS, twocolumn, usenatbib]{mnras}
\usepackage{graphicx}
\usepackage{natbib}
\usepackage{multicol}
\usepackage{txfonts}
\usepackage{ae,aecompl}

\usepackage{accents}
\newcommand{\vardbtilde}[1]{\tilde{\raisebox{0pt}[0.93\height]{$\tilde{#1}$}}}

\def\apj{ApJ}
\def\apjl{ApJL}

\def\aap{A\&A}

\def\prd{PhRvD}

\def\beq#1{\begin{equation}\label{#1}}
\def\eeq{\end{equation}}
\def\beqa#1{\begin{eqnarray}\label{#1}}
\def\eeqa{\end{eqnarray}}
\def\eq#1{eq.~(\ref{#1})}

\def\eqn#1{(\ref{#1})}

\title[Thermalization in hydrostatic atmospheres]{On  thermalization of radiation  in  hydrostatic atmospheres}
\author[M. Gornostaev
] {Mikhail Gornostaev\thanks{E-mail: mgornost@gmail.com}\\
Sternberg Astronomical Institute, Lomonosov Moscow State University, Universitetskij pr. 13, Moscow 119234, Russia
}

\begin{document}

\pagerange{\pageref{firstpage}--\pageref{lastpage}} \pubyear{2022}

\maketitle

\label{firstpage}

\begin{abstract}

The problem of thermalization of radiation within a self-emitting hydrogen isothermal atmosphere is considered  for the case of a hydrostatic profile of the plasma density.
The probabilistic approach to define  the thermalization depth for the photon of a given frequency is used. 
Quantitative conclusions are made on the value of the optical depth at which the radiation can be viewed as thermalized up to a given frequency. The analytical and numerical solutions of the radiative transfer equation confirm the obtained results. The frequency dependencies of the probability of free-free absorption of a photon due to single interaction in a cold nonmagnetized and magnetized plasma are calculated.

\end{abstract}

\begin{keywords}
radiative transfer -- opacity  -- radiation mechanisms: thermal --   stars: neutron -- X-rays: binaries -- magnetic fields
\end{keywords}

\section{Introduction} \label{sec:intro}

The interaction of  radiation with  matter is of an exceptional interest in  many astrophysical researches. 
In the framework of the corresponding problems it is often  useful to introduce  various characteristic length scales describing this interaction and providing  information about the local state of the matter and radiation field. 
Whiles the mean free path  is determined the length between points of two consecutive interaction events, in  media for which  investigating the processes of interaction can be reduced  to the consideration of only  scattering and `true' absorption of photons, the relevant length scale is also the  value of displacement  from the point of the origination of a photon to the point of its destruction due to absorption (the effective mean path,
or the diffusion length).  The latter process takes away the energy from a radiation field and transforms it to the energy of the thermal motion of  particles. Describing such an event one usually says that a photon is thermalized. In this text, however, let us refrain from using the term `thermalization length' instead of the term `effective mean path'.

The state  in which the radiation is fully thermalized corresponds to a source function equal to the Planck function.   
An increase of the diffusion length compared to the mean free path can occur, for example, when the considered point  shifts to the boundary of the medium: then the radiation spectrum at this point, in contrast with the spectrum at the deep, ceases  to be Planckian (at least, to be such within the whole frequency range).

I consider the  problem related with the  formation of a spectrum within the atmosphere  being in  a gravitation field under the conditions when the radiative transfer is determined by the values of the  effective mean path characterizing the photon of a given frequency. 
In  Section~\ref{sec:depth} with the use of the known probabilistic way for defining the thermalization depth (see e.g.~\citealt{1982stat.book.....M}), the expression is derived that describes such an optical depth dependent on the frequency and other parameters at which the radiation at a given frequency is thermalized. 
In Section~\ref{sec:transf} the problem of the spectrum formation is discussed, the analytical and numerical solutions of corresponding radiative transfer equations are obtained and investigated. In Section~\ref{sec:remarks} the comments are adduced concerning the effects that appears in  highly magnetized atmospheres and atmospheres of accreting objects. Section~\ref{sec:concl} contains the conclusions.

\section{Thermalization of radiation at different frequencies} \label{sec:depth}

Consider first a semi-infinite plane-parallel atmosphere consisting of a hydrogen plasma in the absence of a magnetic field.  Let this atmosphere be self-emitting~--- solely its emission is the subject of the investigation,  there are no any external sources irradiating the matter from the outer space.  We now restrict ourselves by the case of the (monohromatic) Thomson scattering supposing that \mbox{$kT\ll mc^2$} and 
\mbox{$h\nu \ll mc^2$}, where $T$ is the electron temperature which is constant within the medium, $k$ is the Boltzmann constant, $m$ is the electron mass, $h$ is the Planck constant and $c$ is the speed of light.  The effect of Comptonization is thus neglected. Free-free processes are taken into account in the frame of the nonrelativistic treatment  in the dipole approximation.

A photon being considered at a given point within the atmosphere can be characterized by the  
probability of free-free absorption in a single interaction 
\beq{eq:pa1}
p=\frac{\sigma_{\rm a}}{\sigma_{\rm T}+\sigma_{\rm a}},
\eeq
where $\sigma_{\rm T}$ is the Thomson cross-section and $\sigma_{\rm a}$ is the free-free absorption cross-section.
The  
expression 
\beq{eq:pa}
p= \frac{\sigma_{\rm a}}{\sigma_{\rm T}}
\eeq
can be applied when the contribution of absorption to the total opacity is relatively small (the medium `is dominated' by scattering).

If the medium is optically thick to the Thomson scattering, the mean number of scatterings  
\beq{eq:Nsc}
N\approx \tau^2,
\eeq
where 
$
\tau=-\sigma_{\rm T} \int n_{\rm e} dz
$
is the Thomson optical depth counted and increased into the atmosphere, with the ray $z$ is oppositely directed   perpendicularly to the boundary which corresponds to the value \mbox{$\tau=0$}, and $n_{\rm e}$ is the electron number density.
Then, as it is known, the solution of the equation (\citealt{1982stat.book.....M,1992hrfm.book.....M,1993ApJ...418..874N})
\beq{eq:Nscpa}
pN=1
\eeq
leads to an answer to the problem about the thermalization depth, i.e., in our case, about the value of the Thomson optical depth $\tau$ corresponding to the point where a photon can be considered as thermalized.

At  frequencies $\nu\gtrsim k T/h$, at which stimulated emission can be ignored, the free-free absorption cross-section reads 
\citep{1979rpa..book.....R, 1992hrfm.book.....M, 1961ApJS....6..167K}
\beq{eq:sigmaff_nm}
\sigma_{\rm a}= \frac{4\pi}{3\sqrt{3}} g_{\rm f\!f} \frac{ e^6 n_{\rm e}}{h c m  \nu^3}\left(\frac{2}{\pi m kT}\right)^{\frac{1}{2}},
\eeq
where $e$ is the elementary charge,  $g_{\rm f\!f}$ is the Gaunt factor.
Taking into account \eqn{eq:pa} and \eqn{eq:Nsc}, let us found the thermalization depth $\tau_{\rm th}$ from equation \eqn{eq:Nscpa}, which is thus written as follows:
\beq{eq:th2}
\frac{\sigma_{\rm a}}{\sigma_{\rm T}} \tau_{\rm th}^2=1.
\eeq

Let  the considered model describe the region of the atmosphere of a gravitating object with  mass $M$ and   radius~$R$.
From the hydrostatic equilibrium equation it is easy to obtain the value of the electron number density $n_{\rm e}$  at the optical depth corresponding to the depth $\tau_{\rm th}$: 
\beq{eq:n_e}
n_{\rm e}=\frac{GM}{R^2}\frac{m_{\rm p} \tau_{\rm th}}{2\sigma_{\rm T} k T},
\eeq
where $G$ is the gravitational constant and $m_{\rm p}$ is the proton mass. 

Supposing that $g_{\rm f\!f} = 1$ and substituting \eqn{eq:n_e} in \eqn{eq:sigmaff_nm}, and after that \eqn{eq:sigmaff_nm} in \eqn{eq:th2},
one can obtain under the made approximations the depth \citep{nkr2020}
\beq{eq:tau_th}
\tau_{\rm th} \approx \frac{h\nu}{1~{\rm {keV}}} \left( \frac{k T}{1.5~{\rm {keV}}} \right)^{\frac{1}{2}}
\left(\frac{M}{1.5M_\odot}\right)^{-\frac{1}{3}}\left(\frac{R}{10^6~{\rm {cm}}}\right)^{\frac{2}{3}}.
\eeq
The quantity $\tau_{\rm th}$ can be called  the depth of `partial' thermalization. It indicates  the optical depth 
$\tau$ at which the radiation can be considered as thermalized up to the frequency~$\nu$.

 Equation \eqn{eq:Nscpa}, as well as some of sort of it used in the theory, is only a useful criterion for separating  thermalized photons from other since  there is no any condition describing this transition exactly. 
 This is, however, a known thing that concerns characteristic frequencies and optical depths and is manifested in the radiative transfer problems. 
 The value  $\tau_{\rm th}=1$ points out the frequency (with an error of less than a  per~cent), at which 
 $\left.\sigma_{\rm a}\right|_{\tau=1}=\sigma_{\rm T}$ at the given parameters (and under the made assumptions).
  As the absorption probability tends to zero along with the frequency, the quantity $\tau_{\rm th}$ formally tends to infinity.
 One can nevertheless believe  that the extent of thermalization of the radiation corresponds to LTE, say, the spectral
 radiation energy density $u_\nu$ becomes approximately equal 
 to the frequency-integrated value,
 \beq{}
 \int_0^{\nu(\tau_{\rm th})} u_\nu(\tau_{\rm th}) d\nu \approx u,
 \eeq 
 where $\nu(\tau_{\rm th})$ is found from \eqn{eq:tau_th}.
At a known  temperature, 
 using the Planck distribution it is possible to estimate  the frequency range in which almost all photons at the corresponding  depth are thermalized.

\section{Spectrum formation} \label{sec:transf}



The solutions of the radiative transfer equation
can indicate 
the evolution of the spectrum as a whole along $\tau$. 

In the first version of the contribution, it was noted that the number of scatterings in the consideration above is described by the optical depth written through the Thomson opacity. This should be kept in mind. 
It would be possible, under this assumption,  to consider only some refinements to the formulas obtained above. One can immediately see thus, for instance, that the linearity of  \eqn{eq:tau_th} in frequency  can be broken (at certain frequencies) due to the
distinction of the Gaunt factor from unity \citep{1961ApJS....6..167K} and due to stimulated emission, with taking into account of which the absorption cross-section reads \citep{1979rpa..book.....R, 1992hrfm.book.....M}
\beq{eq:abscorr}
\sigma_{\rm a}'=\sigma_{\rm a}\left(1-{\rm e}^{-\frac{h\nu}{kT}}\right).
\eeq
These changes in cross-section, however, should not affect the obtained conclusion significantly.
An important  role in obtaining solution \eqn{eq:tau_th} is played by the chosen electron number density profile \eqn{eq:n_e}, which is determined at a fixed temperature only by the column density of the matter (e.g. \citealt{1969AZh....46..225Z,  1993ApJ...418..874N,1994ApJ...426L..35T}), with the pressure at the boundary is equaled to zero.

The effect of changing the frequency of photons due to scatterings on  the spectrum formation  
does to be somewhat inaccurate the obtained result in any case (see e.g. \citealt*{pavlov1989} for the review), especially if the temperature is too high. The corresponding situations are often studied numerically. In the considered circumstances the effect of the redistribution over the angles in the radiation-diffusion regime should also not be very important (\citealt{1982stat.book.....M}).

The use in \eq{eq:Nscpa} of the expression
$N\approx{\rm max}\{\tau^2, \tau\}$
or 
$N\approx\tau^2+\tau$
for the mean number of scatterings (\mbox{\citealt{1979rpa..book.....R}}) does not significantly change the results for the depth $\tau_{\rm th}$.

Note that for a homogeneous medium from \eqn{eq:th2} one has \mbox{$\tau_{\rm th} \propto \nu^{3/2}$}.

\subsection{Analytical solutions}

In the case of  coherent scattering, the radiative transfer with taking into account absorption 
under the conditions of the hydrostatic atmosphere with the predominance of scattering 
is described by \cite{1969AZh....46..225Z},
\linebreak appendix~1. The radiative transfer equation
\beq{eq:transfm}
\mu \frac{\partial I_\nu}{\partial\tau_\nu}=  I_\nu - S_\nu 
\eeq
has being considered, where  $I_\nu$ is the radiation intensity, $\mu$ is  the cosine of the polar angle which is originated by the given direction,  the optical depth $\tau_\nu$ is introduced by means of \mbox{${\rm d}\tau_\nu=-(\sigma_{\rm T}+\sigma'_{\rm a})n_{\rm e}{\rm d} z$}, and
\beq{eq:Snu}
S_\nu=\frac{\sigma_{\rm T} J_\nu+\sigma'_{\rm a} B_\nu}{\sigma_{\rm T}+\sigma'_{\rm a}}
\eeq
is the source function, with  $J_\nu=\frac{c}{4\pi}u_\nu$ is the mean intensity, $B_\nu$ is the Planck function.
In accordance with the known result of the work of \cite{1969AZh....46..225Z}, the solution (in the Eddington approximation) for $J_\nu$  within the medium  can be expressed by means of the  Airy functions \citep{abramovitz}. 
Thus, solving the transfer equation 
\beq{eq:transf}
\frac{\partial^2 J_\nu}{\partial\tau^2}= \frac{3\sigma'_{\rm a}}{\sigma_{\rm T}}(J_\nu-B_\nu)
\eeq 
with taking into account the boundedness of the solution in the deep  and with the use of the free-escape (Marshak) boundary condition
\beq{}
\frac{1}{\sqrt{3}}\frac{\partial J_\nu}{\partial\tau}=J_\nu
\eeq
one can obtain
\beq{eq:J}
J_\nu=B_\nu  +  C Ai(\xi), 
\eeq
where     \mbox{$\xi=\tau C_0^{1/3}$}, with 
\mbox{$C_0= 3\sigma'_{\rm a}/(\sigma_{\rm T}\tau)$}, and
\beq{}
C=-\frac{3^{5/6}  \Gamma(1/3) \Gamma(2/3) B_\nu }{3^{1/6}\Gamma(1/3)+C_0^{1/3}\Gamma(2/3)}.  
\eeq

\begin{figure}
 	\begin{center}
  \includegraphics[width=0.48\textwidth]{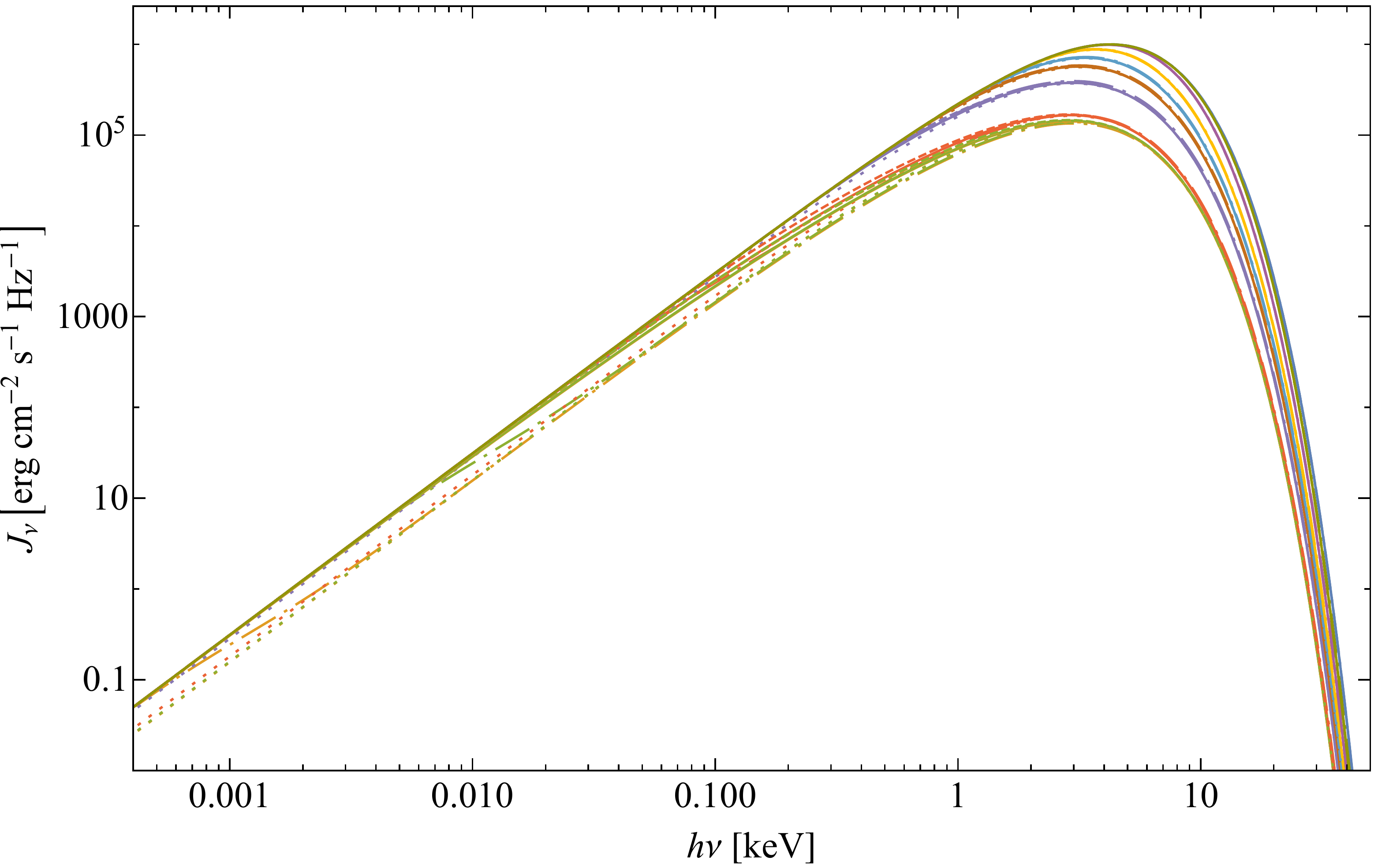}
  		\caption{Solutions  \eqn{eq:J} (solid), \eqn{eq:Jabs} (dashed), \eqn{eq:Jabs2} (dotted), and solution obtained numerically (dot-dashed lines) for $J_\nu$ are shown for different values of $\tau$: 0.001 (light brown, joins with the following and is heavily distinguished), 0.01 (light green), 0.1 (light red), 1 (purple), 2 (brown), 3 (blue), 5 (yellow), 10 (dark purple), 20 (green) and~$\infty$ (dark blue). The problem parameters: \mbox{$T=1.5~{\rm keV}$}, \mbox{$M=1.5M_\odot$} and \mbox{$R=10^6~{\rm cm}$}.}
  \label{fig:Jnu}
	\end{center}
 \end{figure}

The results for different values of $\tau$ are shown in Fig.~\ref{fig:Jnu}, an infinite value corresponds to the Planck function. The frequency of the right boundary of the range in which the spectra coincide with the Planck function actually changes with $\tau$ according to a law that seems to be linear. The specific values of the thermalization depth, however, slightly differ from those given by \eq{eq:tau_th}.
The approximate fits (e.g. corresponding to Fig.~\ref{fig:Jnu})  give  
\beq{}
\tau^*_{\rm th}(\nu, T, M, R)\approx 2 \tau_{\rm th}(\nu, T, M, R),
\eeq
where $\tau^*_{\rm th}$ denotes the thermalization depth found from the calculated spectra, $\tau_{\rm th}$ is found from \eqn{eq:tau_th}. 
This difference is not due to using in  \eq{eq:transf} the simple expression \eqn{eq:pa} for the absorption probability.
The choice of the certain  free-escape boundary condition  does not influence  on the solution significantly.

With increasing temperature, the ratio of the total radiation energy density in the near-surface layers (\mbox{$\tau \sim 1$}) to the value in the completely thermalized ones decreases. At relatively low temperatures, thermalization takes place near the boundary, so that the atmosphere shines like a black body. The ratio \mbox{$u|_{\tau=1}/(aT^4)$}, where $a$ is the radiation constant, reaches  unity 
at \mbox{$T \sim 0.1~{\rm keV}$} 
(when $M$ and $R$ are the same as in calculations above).

Now consider the radiative transfer equation written with taking into account \eqn{eq:pa1} in more general form (\mbox{\citealt{1979rpa..book.....R}})  
\beq{eq:transf2}
\frac{\partial^2  J_\nu}{\partial\tau_\nu^2}= \frac{3\sigma'_{\rm a}}{\sigma_{\rm T}+\sigma'_{\rm a}}(J_\nu-B_\nu).
\eeq
Rewriting \eq{eq:transf2} as
\beq{eq:transf21}
\frac{\partial^2 J_\nu}{\partial\tau^2}=\frac{3\sigma'_{\rm a}} {\sigma_{\rm T}} \left(1+\frac{\sigma'_{\rm a}}{\sigma_{\rm T}}\right)
(J_\nu-B_\nu)
\eeq
and solving \eq{eq:transf21} with the use of the same conditions at the boundary and infinity as in the previous case, one can obtain
\beq{eq:Jabs}
J_\nu=B_\nu+\tilde C D_{\frac{3\sqrt{3}-4C_0}{8C_0}}\left(\frac{3+2C_0\tau}{3^{1/4}\sqrt{2C_0}}\right),
\eeq
where the second term is expressed by means of the parabolic cylinder function \citep{abramovitz}, and
\beq{}
\tilde C = -\frac{6 B_\nu}{3^{1/4}2\sqrt{2C_0}D_{1+\frac{3\sqrt{3}-4C_0}{8C_0}}\left(\frac{3^{3/4}}{\sqrt{2C_0}}\right)
+3D_{\frac{3\sqrt{3}-4C_0}{8C_0}}\left(\frac{3^{3/4}}{\sqrt{2C_0}}\right)}.
\eeq

The solution is shown in Fig.~\ref{fig:Jnu}. It is clear that in the high-frequency range 
the spectra are easily represented by  scattering-dominated solution~\eqn{eq:J}, this was used when constructing the figure. 
The sense of solution \eqn{eq:Jabs} is not very big because the difference between \eq{eq:transf} and \eq{eq:transf2}  consists in appearing in right-hand side of \eq{eq:transf21} of the quadratic term in $\tau$ which rapidly decreases in the near-surface region (therefore, at low frequencies the significant contrast between two solution absents).

Looking at these solutions, one might think that, as  $\tau$ decreases, starting from its certain  value, the dependence of $J_\nu$ on $\tau$ disappears (when it exists at all), and the range in which the mean intensity coincides with the Planck function ceases to narrow. The frequency corresponding to the right boundary of this range is the same for both solutions in order of magnitude. 
For example, at the temperature of~1.5~keV (for the mentioned parameters of the neutron star) its value lies between $10^{-3}$~keV and 
$10^{-2}$~keV for solution \eqn{eq:J} and approximately corresponds to energy $10^{-2}$~keV   for  solution \eqn{eq:Jabs},
the values of $\tau$ do not exceed $\sim 10^{-2}$. Such a behavior of spectra, as it will be seen below from the numerical solutions, is only a feature  of the considered approximation, and, actually, the frequency up to which $J_\nu=B_\nu$ at a given $\tau$ moves to the side of 
low frequencies with decreasing $\tau$ once leaving the X-ray range.

Note that considering the equation
\beq{eq:transfw}
\frac{\partial^2 J_\nu}{\partial\tau^2}= \frac{3\sigma'_{\rm a}}{\sigma_{\rm T}+\sigma'_{\rm a}}(J_\nu-B_\nu)
\eeq
and using again the boundary conditions as in previous cases lead to the solution 
\beq{eq:Jabs2}
J_\nu=B_\nu+\vardbtilde{C} W_{\frac{3\sqrt{3}}{2C_0},-\frac{1}{2}}\left( 2\sqrt{3}\left(\tau+\frac{3}{C_0}\right) \right)
\eeq
written through the Whittaker function \citep{abramovitz}, where
\beq{}
\vardbtilde{C} = -\frac{6 B_\nu}{\frac{2}{\sqrt{3}}C_0W_{1+\frac{3\sqrt{3}}{2C_0},-\frac{1}{2}}\left(\frac{6\sqrt{3}}{C_0}\right) +
3W_{\frac{3\sqrt{3}}{2C_0},-\frac{1}{2}}\left(\frac{6\sqrt{3}}{C_0}\right)}.
\eeq

This solution (Fig.~\ref{fig:Jnu}) does not take into account changing the quantity ${\partial \tau_\nu}/{\partial z}$ 
with frequency (arising due to free-free absorption) and  does not work well for small $\tau<1$, at low frequencies. 
For $\tau\rightarrow 0$, however, when the indicated approximation is applicable, and the frequency at which the spectrum merges with the blackbody one is sufficiently low, this solution coincides with the minimum values of $J_\nu$ calculated due to numerical solving \eq{eq:transfm} (see Fig.~\ref{fig:Jnu} and description of the numerical solution below). 
Within the framework of the problem under consideration it is relatively easy to obtain mentioned numerical solution that describes the intensity as a function of the angle, 
therefore solution \eqn{eq:Jabs2} should hardly be given significant importance.

Since (now we do not care which expression is used to calculate $\sigma_{\rm a}$)
\beq{}
{\rm d}\tau_\nu=\left(\frac{\sigma_{\rm a}}{\sigma_{\rm T}}+1\right){\rm d}\tau,
\eeq
one can write that, in the case of the considered density profile \eqn{eq:n_e},
\beq{}
\tau_\nu=\frac{C_1}{2\sigma_{\rm T}}\tau^2+\tau,
\eeq
where $C_1=\sigma_{\rm a}/\tau$. 
Then, if stimulated emission is neglected, the frequency $\nu_0$ fulfilled the equality 
\beq{}
\tau_\nu(\tau)=1
\eeq
at given $\tau<1$  reads
\beq{eq:nu_0}
\nu_0(\tau)=\left(\frac{C_2}{2\sigma_{\rm T}(1-\tau)}\right)^\frac{1}{3}\tau^\frac{2}{3},
\eeq
where $C_2=\sigma_{\rm a}\nu^3/\tau$ (the medium becomes to be optically thin at higher frequencies). For several values of the temperature, the dependencies  are shown in Fig.~\ref{fig:nu0}.





\subsection{Numerical calculations}

Equation \eqn{eq:transfm} can be considered numerically without resorting to the diffusion approximation.  
It is useful to do this exercise in order to see the behavior of the  solutions for the intensity depending on the direction, casually getting rid of the restrictions on the optical thickness on which the considered point is located.

The distributions of intensity are calculated  due to solving equation \eqn{eq:transfm} by Feautrier method \citep{feautrier, 1982stat.book.....M}. The corresponding numerical scheme is based on the well-known tridiagonal matrix algorithm \citep{gelfand}.
The Feautrier equations for the functions $u=\left(I_\nu(\mu)+I_\nu(-\mu)\right)/2$ and $v=\left(I_\nu(\mu)-I_\nu(-\mu)\right)/2$  read
\beq{eq:F1}
\mu\frac{\partial v}{\partial \tau_\nu}=u-S_\nu,
\eeq
\beq{eq:F2}
\mu\frac{\partial u}{\partial \tau_\nu}=v,
\eeq
from which the second-order equation for the function $u$ follows: 
\beq{eq:F3}
\mu^2\frac{\partial^2 u}{\partial \tau_\nu^2}=u-S_\nu.
\eeq
The   top boundary condition is the condition of the free escape of the radiation:
\beq{eq:bctop}
\mu\left.\frac{\partial u}{\partial \tau_\nu}\right|_0=u(0).
\eeq
The bottom one is the Planckian spectrum:
\beq{eq:bcbot}
u+\mu\frac{\partial u}{\partial \tau_\nu}=B_\nu,~~\tau_\nu\rightarrow\infty.
\eeq
When the bottom boundary condition is specified at a sufficient depth, the equality $u=B_\nu$ is also valid (assuming $v=0$)~-- this variant, as is easy to see, does not distort the considered solution.

The choice of the cosines of the angles is realized using Gauss quadratures with four nodes within the interval (0,\,1), which are the roots of the Legendre polynomials of the appropriate degree. A uniform grid in $\tau$ is used. The code is written in C, the algorithm  quite corresponds to the description of \cite{1982stat.book.....M}. The boundary conditions are implemented with the second order of accuracy. An iterative process is constructed to find the source function (the mean intensity) at each iteration, the algorithm for solving the moment equations (see also \citealt{1988ApJ...324.1001B}) is omitted for now.

In Fig. \ref{fig:Jnu} dependencies for the mean intensity are shown found from the solutions of \eq{eq:transfm}.
For small $\tau$, the frequency was   found above at which the transition occurs to the region where $\tau_\nu < 1$ and the Eddington approximation does not work well. In this range and for the lower frequencies a numerical consideration is important. At sufficiently high frequencies, which are of main interest in the problem, all solutions are in perfect agreement.
Nevertheless, the solution of \eq{eq:transfm}  shows that even at $\tau\ll 1$, where the spectrum is not completely described by analytical solutions, the thermalization depth is linear in frequency.

\begin{figure}
 	\begin{center}
  \includegraphics[width=0.48\textwidth]{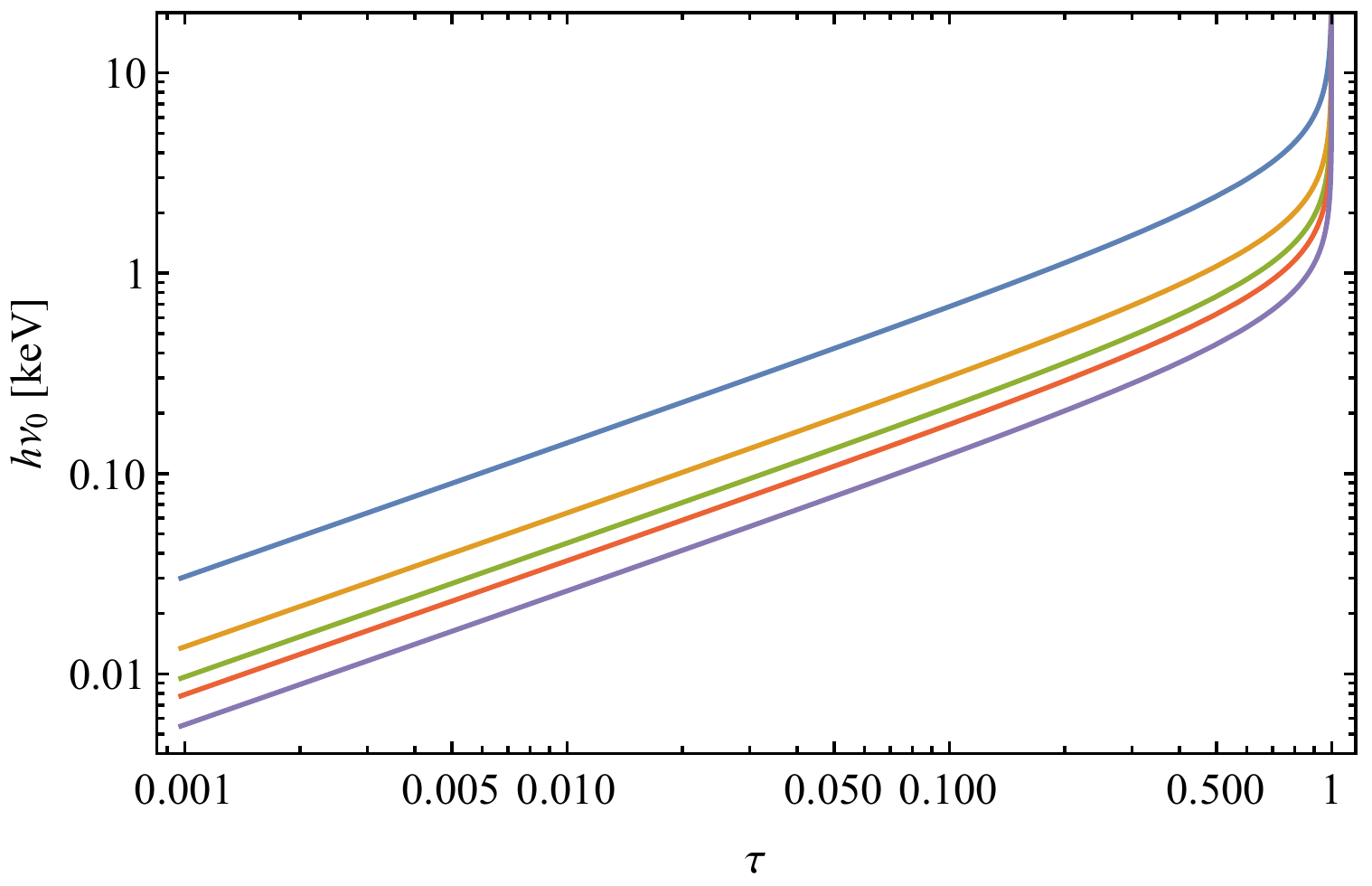}
  		\caption{The photon energy $h\nu_0$ as a function of $\tau$ plotted for different values of temperature: (from top to bottom) 0.1, 0.5, 1, 1.5, 3 keV.}
  \label{fig:nu0}%
	\end{center}
 \end{figure}

It is clear that since $\tau_\nu(\tau) > \tau$ is always the case, at $\tau\rightarrow 1-0$ the range where the diffusion approximation does not work very well ceases to exist, $\nu_0$ tends to infinity (Fig.~\ref{fig:nu0}). For $\tau \lesssim 1$, the numerical solutions  go slightly above than solution \eqn{eq:Jabs2}, without merging with the latter.

\begin{figure*}
 	\begin{center}
     \includegraphics[width=0.32\textwidth]{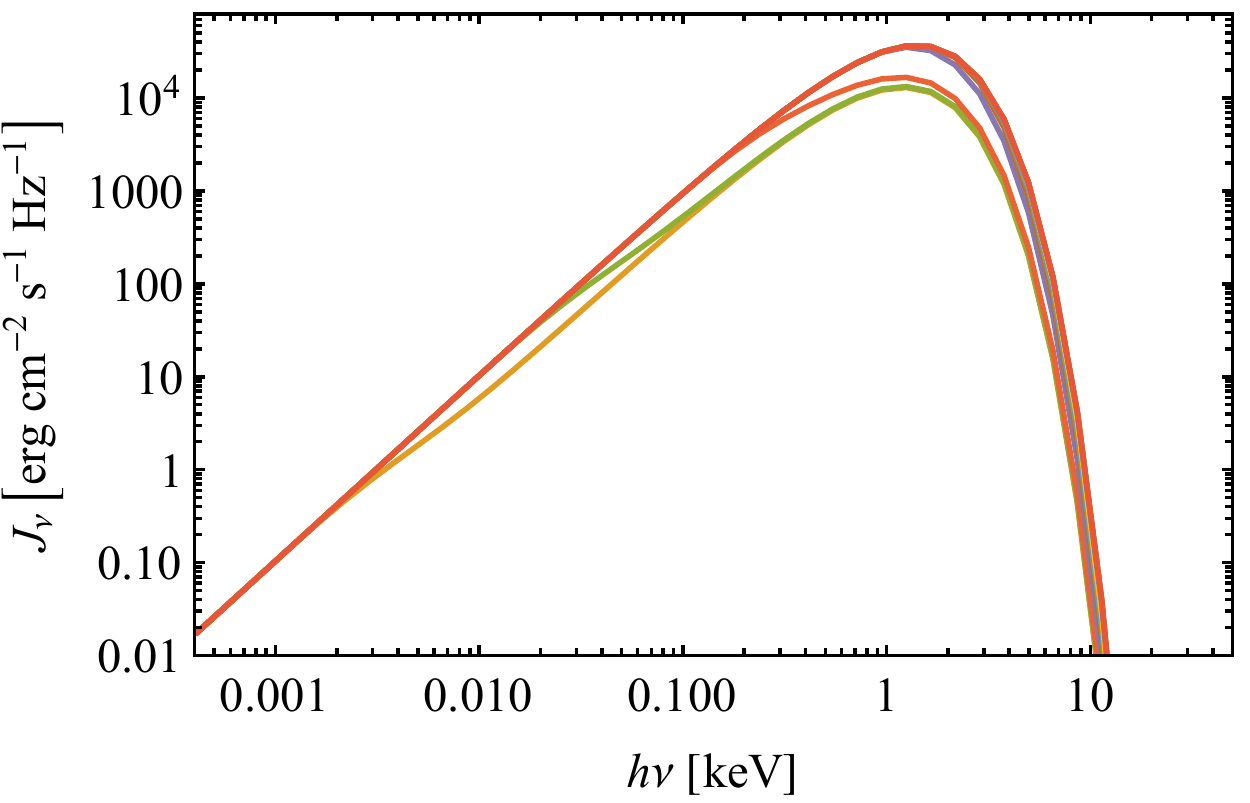}                                
\hfill    
    \includegraphics[width=0.32\textwidth]{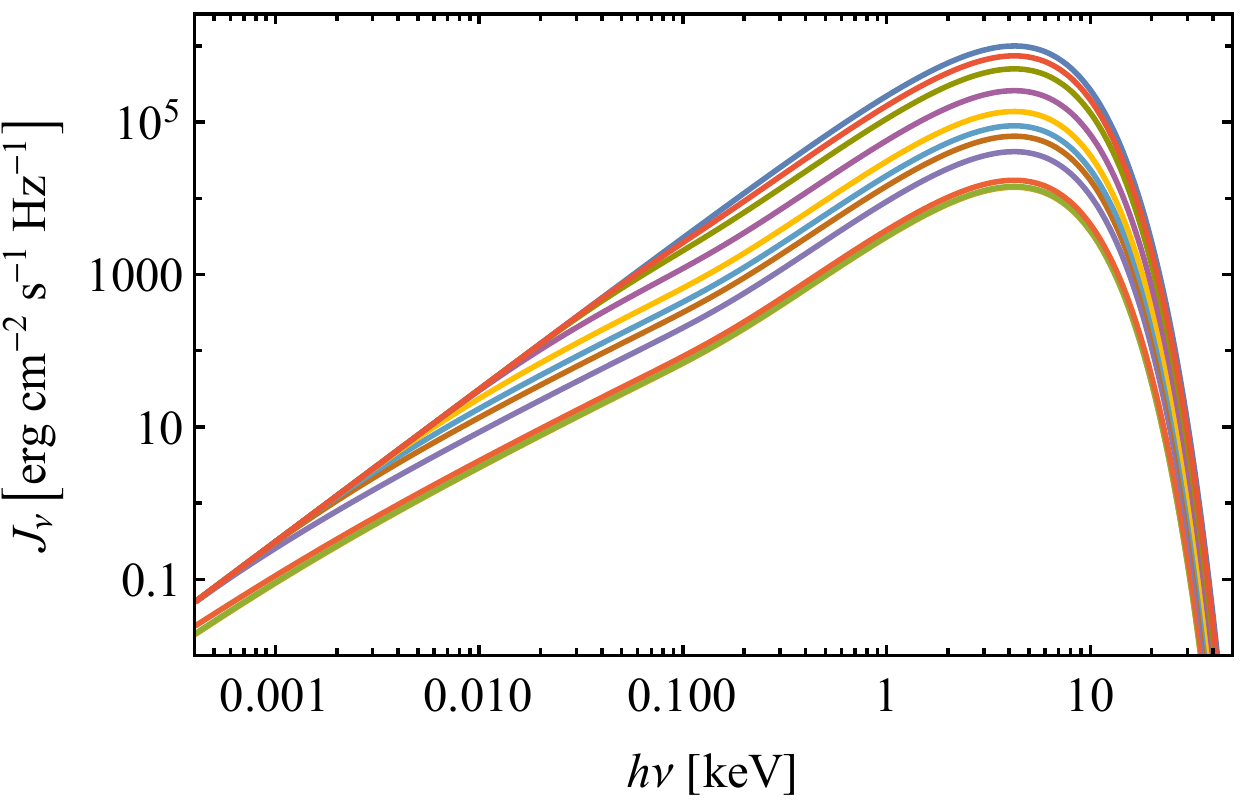}
\hfill
      \includegraphics[width=0.32\textwidth]{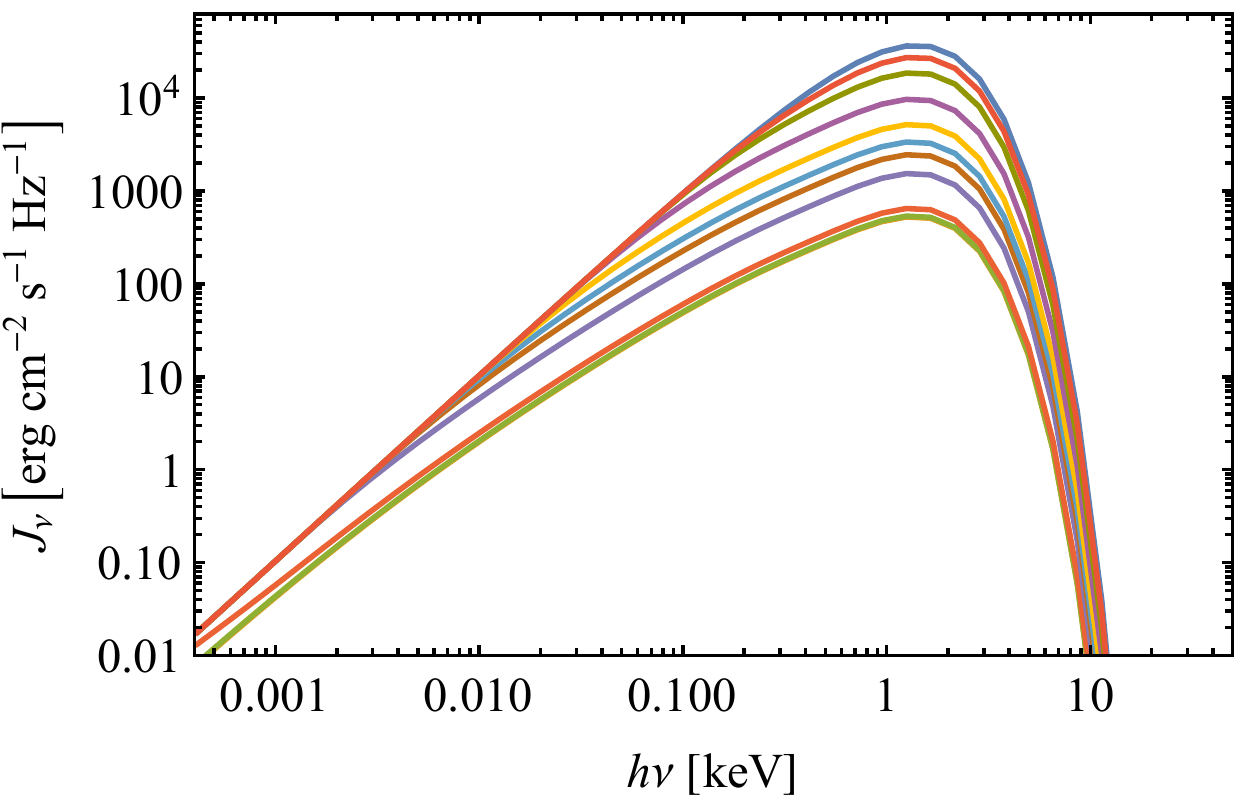}
  \\
     (a) \hspace{155pt} (b) \hspace{155pt} (c)  
  \\  

  		\caption{Solutions for $J_\nu$ obtained in consequence of numerical solving \eq{eq:transfm}  are shown for different $\tau$: 0.001 (light brown, heavily distinguished), 0.01 (light green), 0.1 (light red), 1 (purple), 2 (brown), 3 (blue), 5 (yellow), 10 (dark purple), 20 (green), 30 (red) and~$\infty$ (dark blue). The problem parameters:  \mbox{$T=0.5~{\rm keV}$}, \mbox{$M=1.5M_\odot$} and \mbox{$R=10^6~{\rm cm}$} (a), 
  		 \mbox{$T=1.5~{\rm keV}$}, \mbox{$M=0.6M_\odot$} and \mbox{$R=7\cdot 10^8~{\rm cm}$} (b), \mbox{$T=0.5~{\rm keV}$}, \mbox{$M=0.6M_\odot$} and \mbox{$R=7\cdot 10^8~{\rm cm}$} (c).   }
  \label{fig:Jnudiff}
	\end{center}
 \end{figure*}

Fig.~\ref{fig:Jnudiff} illustrates the dependence on temperature and parameters of a gravitating object (the typical mass and radius of a white dwarf are taken as an example for comparison). In a weaker gravitational field, obviously, for a given $\tau$, the same photon destruction probability  as in a stronger field is achieved  at lower frequencies.

\begin{figure*}
 	\begin{center}
            \includegraphics[width=0.23\textwidth]{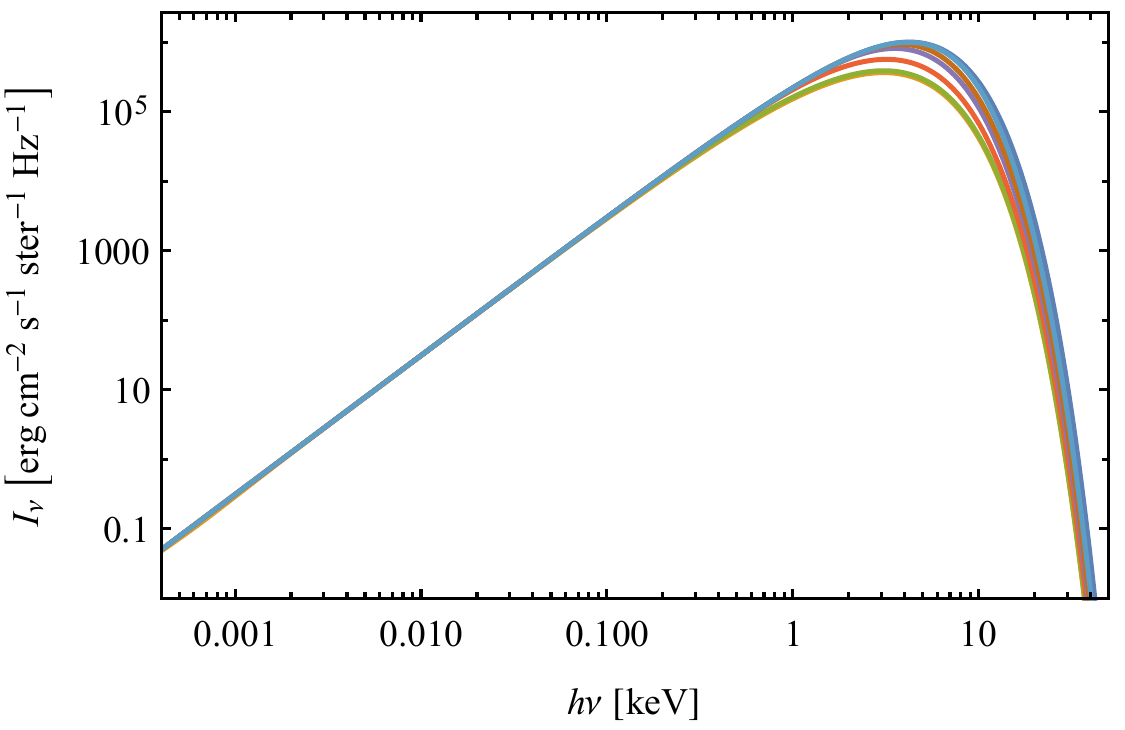}
  \hfill
                                 \includegraphics[width=0.23\textwidth]{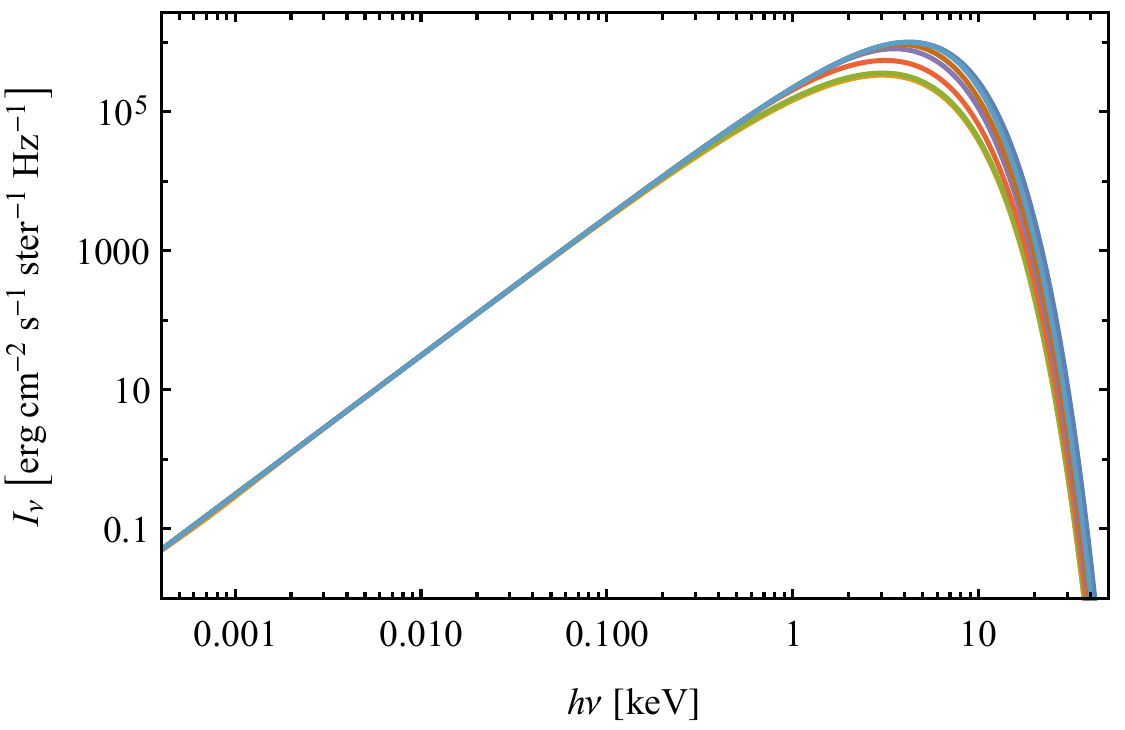}
  \hfill                               
                                   \includegraphics[width=0.23\textwidth]{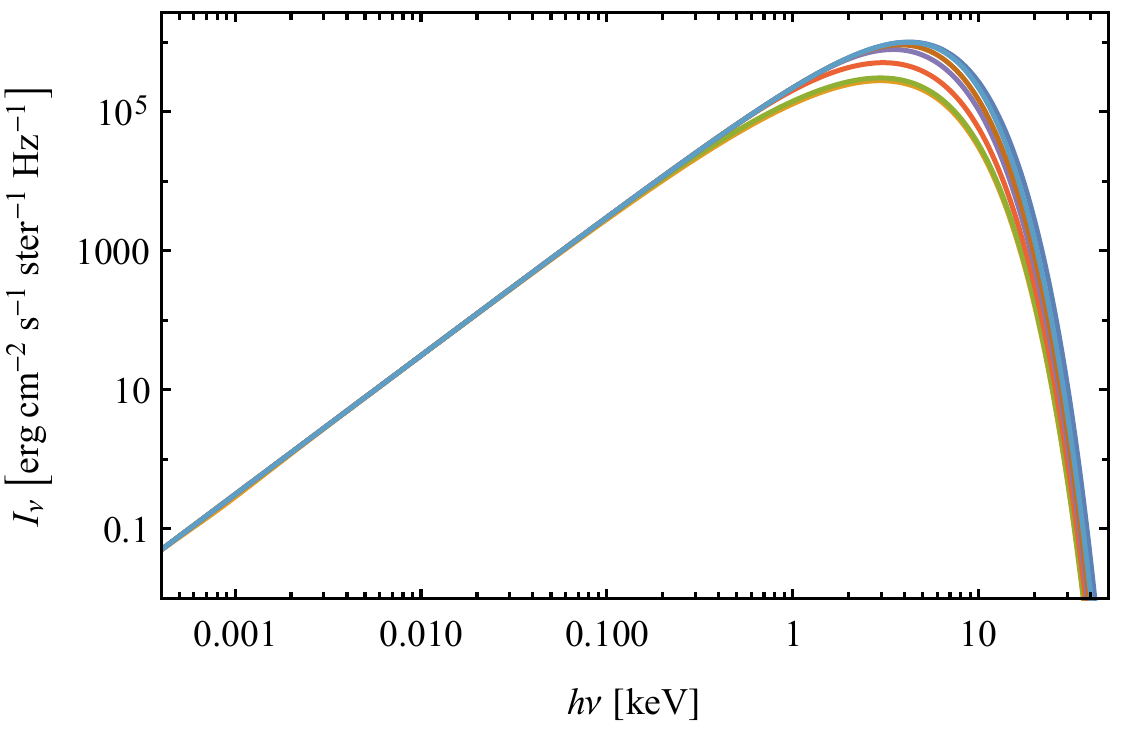}
  \hfill
                                 \includegraphics[width=0.23\textwidth]{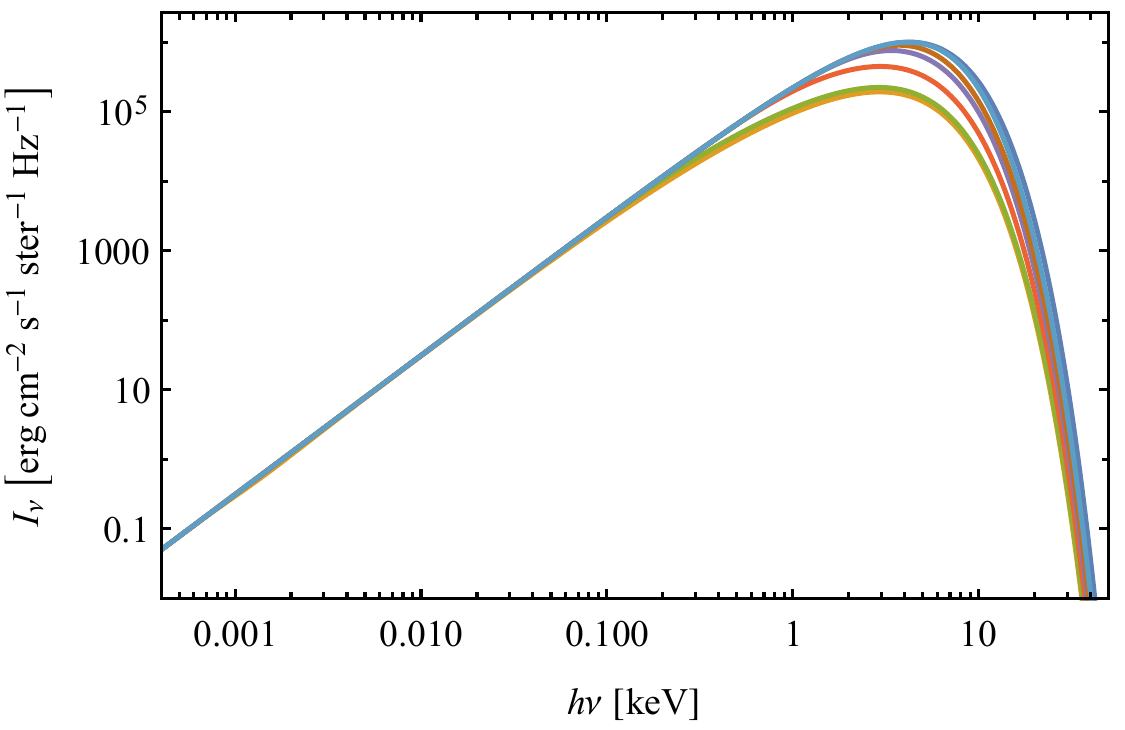}
   \\
        \includegraphics[width=0.23\textwidth]{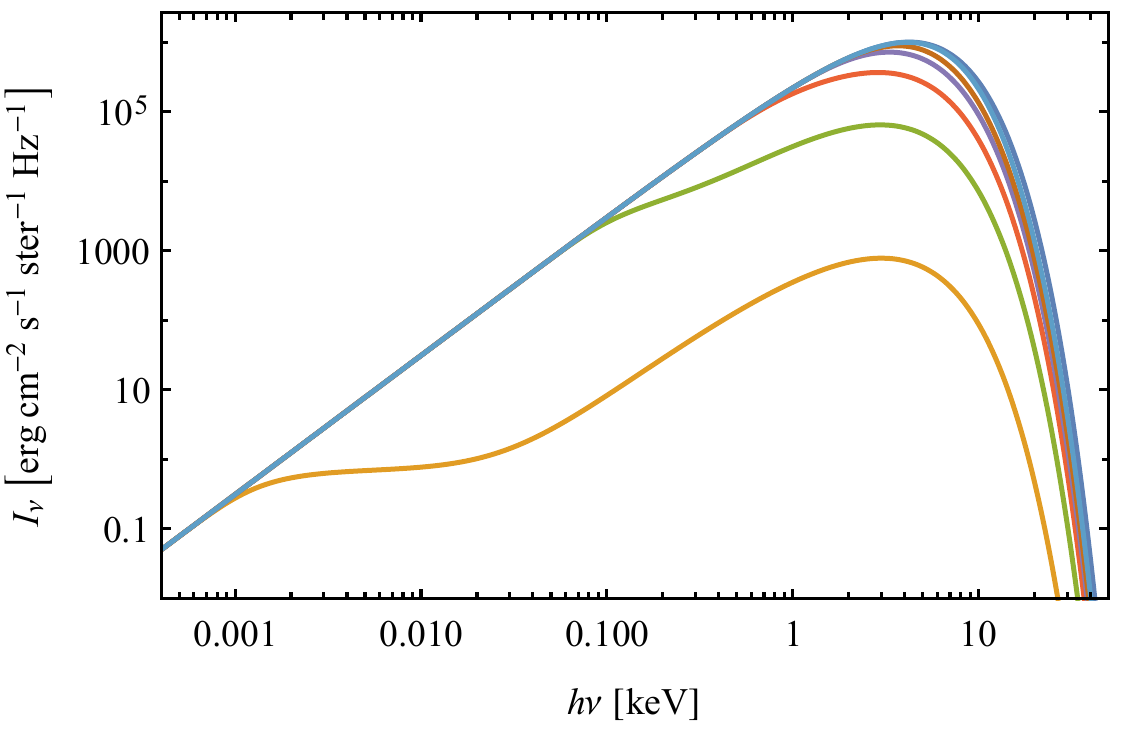}
  \hfill
                                 \includegraphics[width=0.23\textwidth]{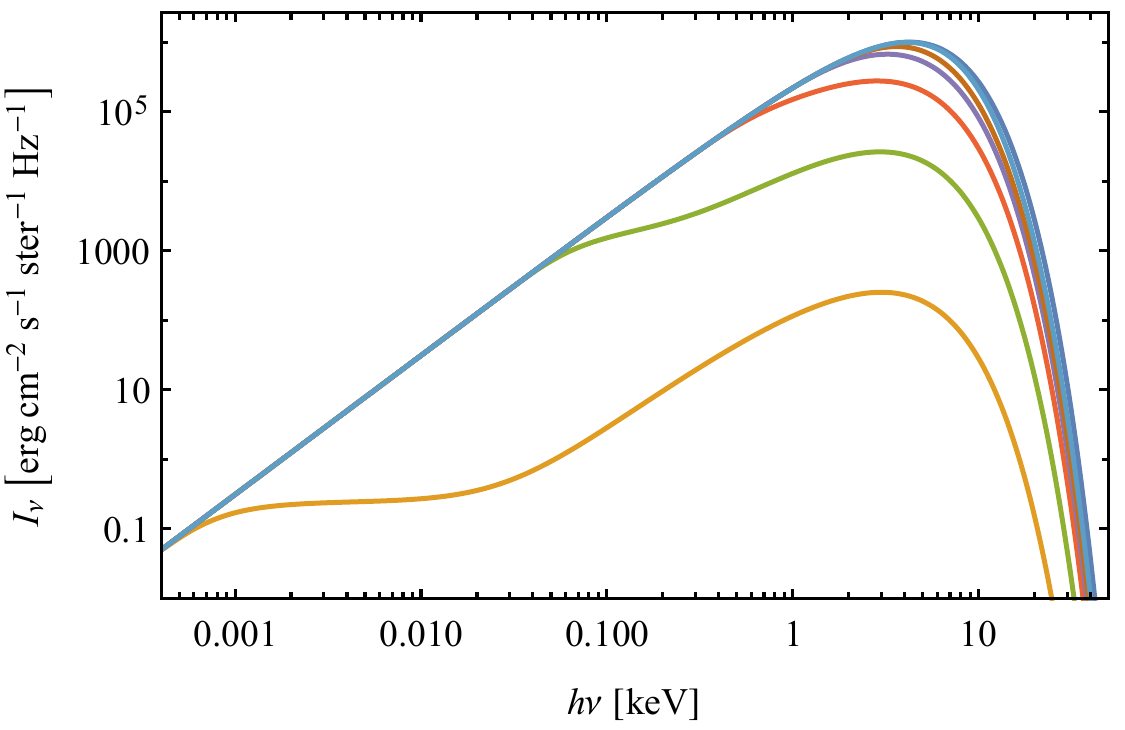}
  \hfill                               
                                   \includegraphics[width=0.23\textwidth]{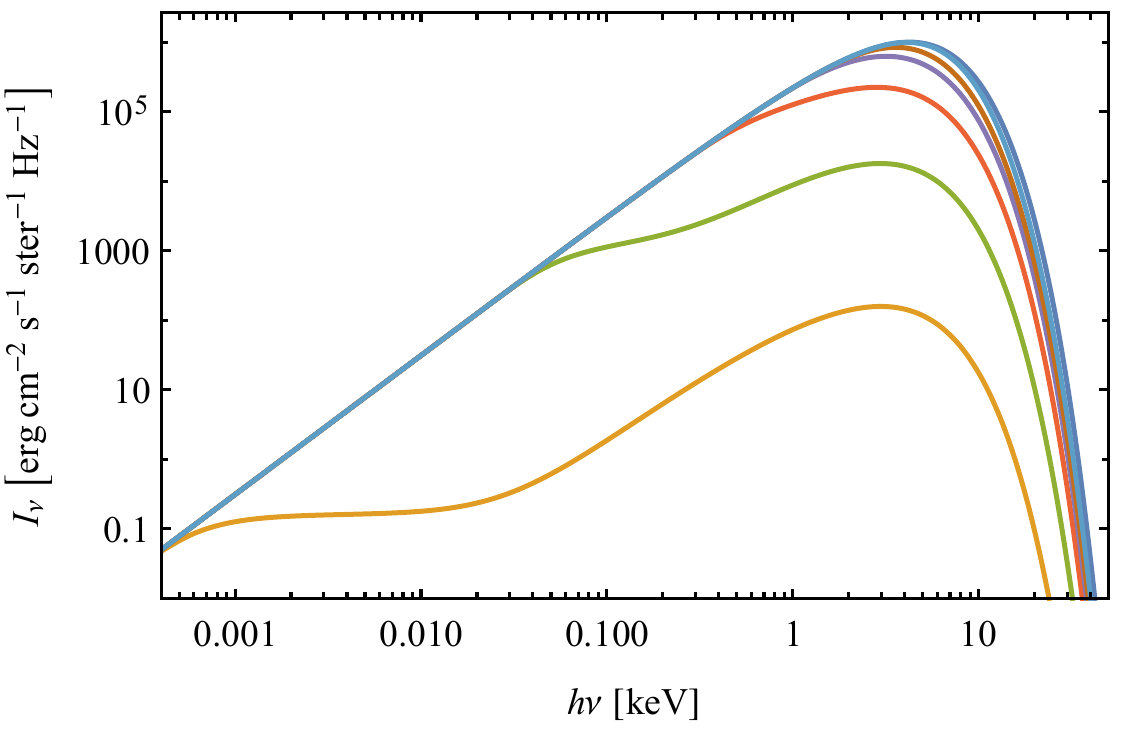}
  \hfill
                                 \includegraphics[width=0.23\textwidth]{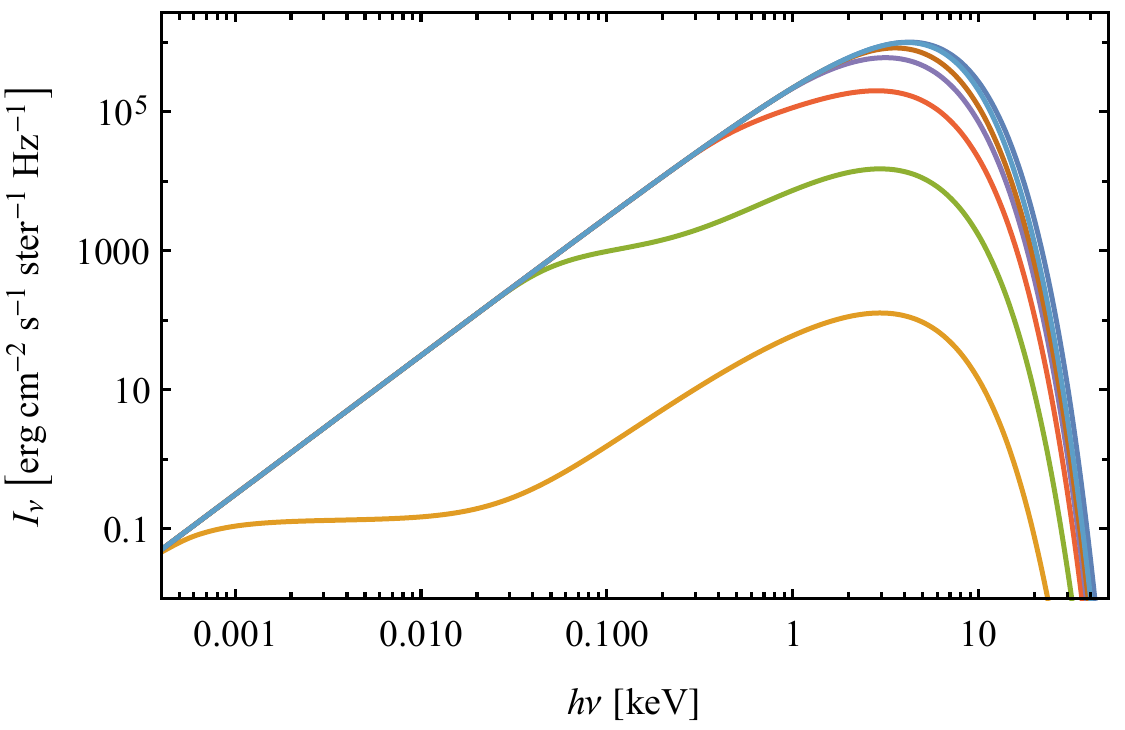}
    \\
   (a) 
  \\
  \vspace{10pt}
            \includegraphics[width=0.23\textwidth]{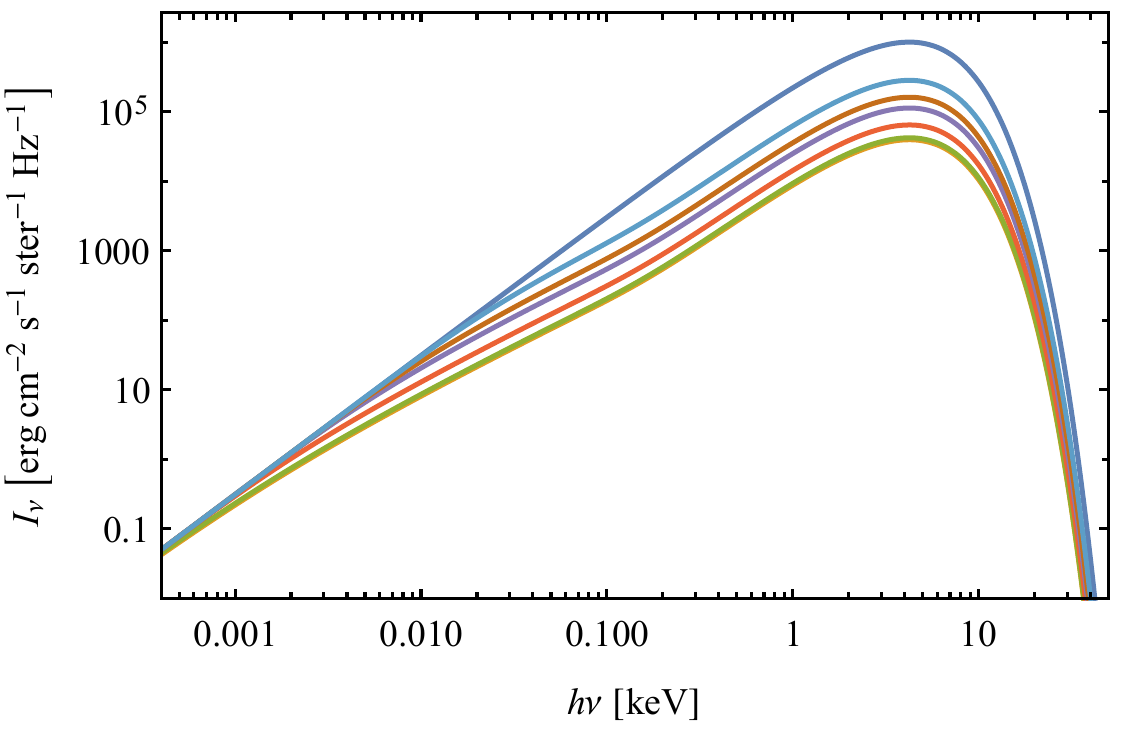}
  \hfill
                                 \includegraphics[width=0.23\textwidth]{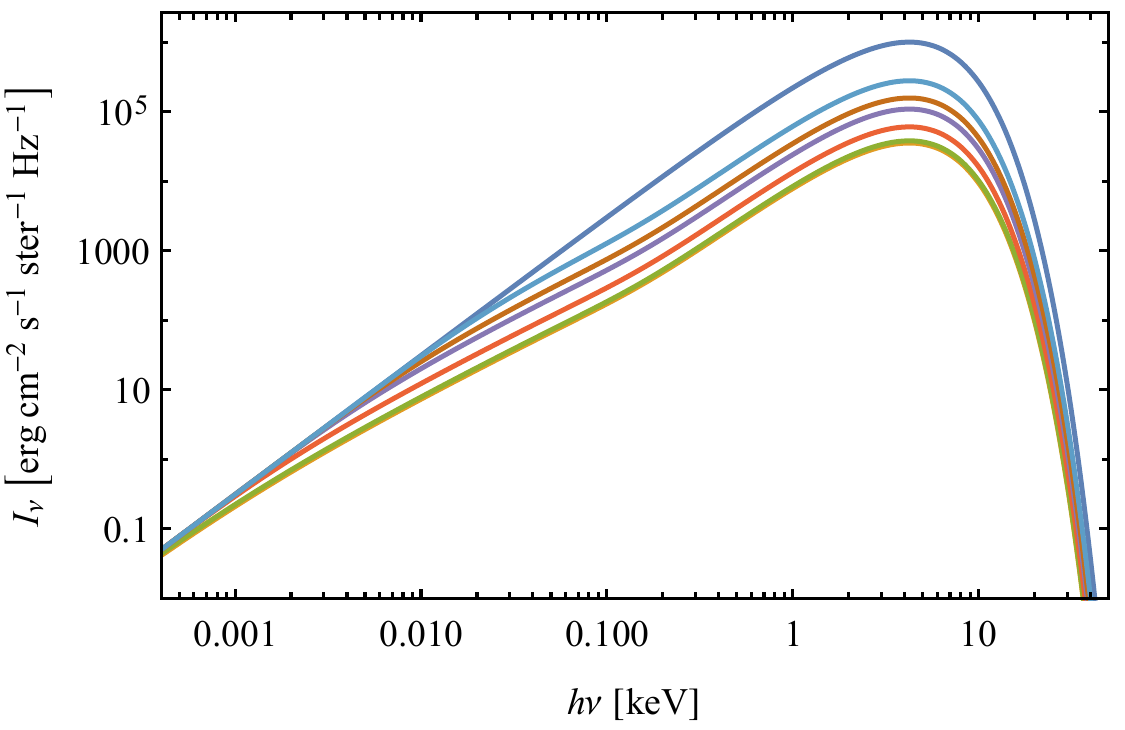}
  \hfill                               
                                   \includegraphics[width=0.23\textwidth]{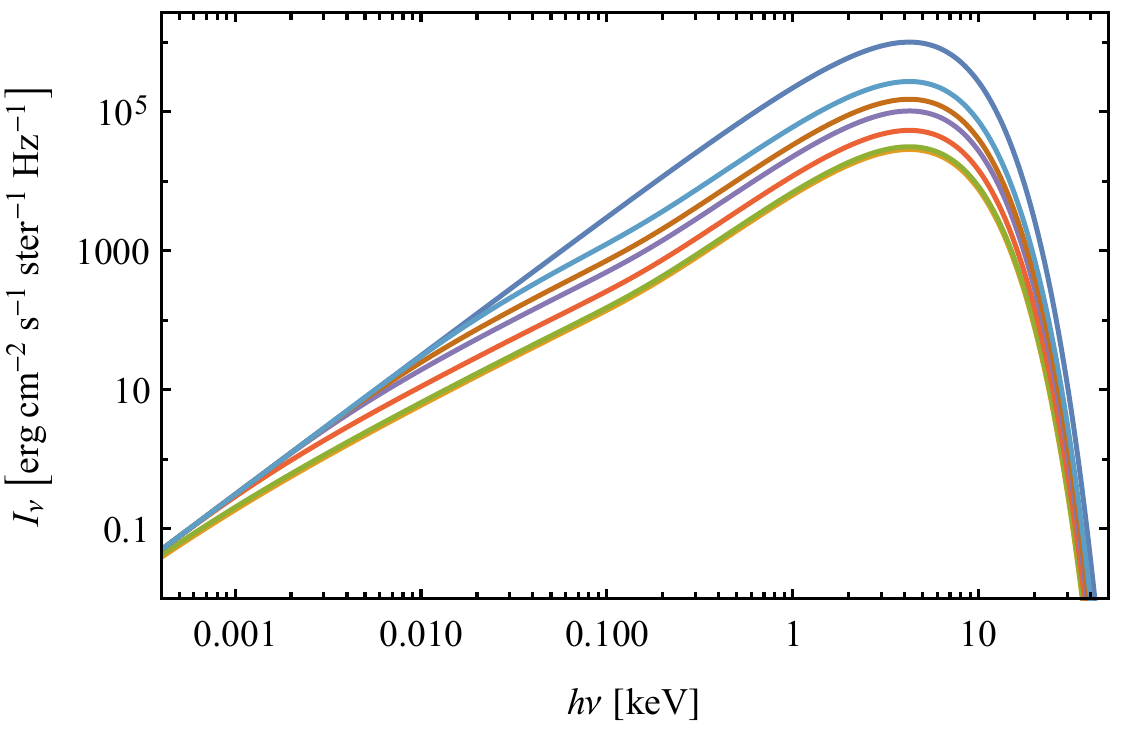}
  \hfill
                                 \includegraphics[width=0.23\textwidth]{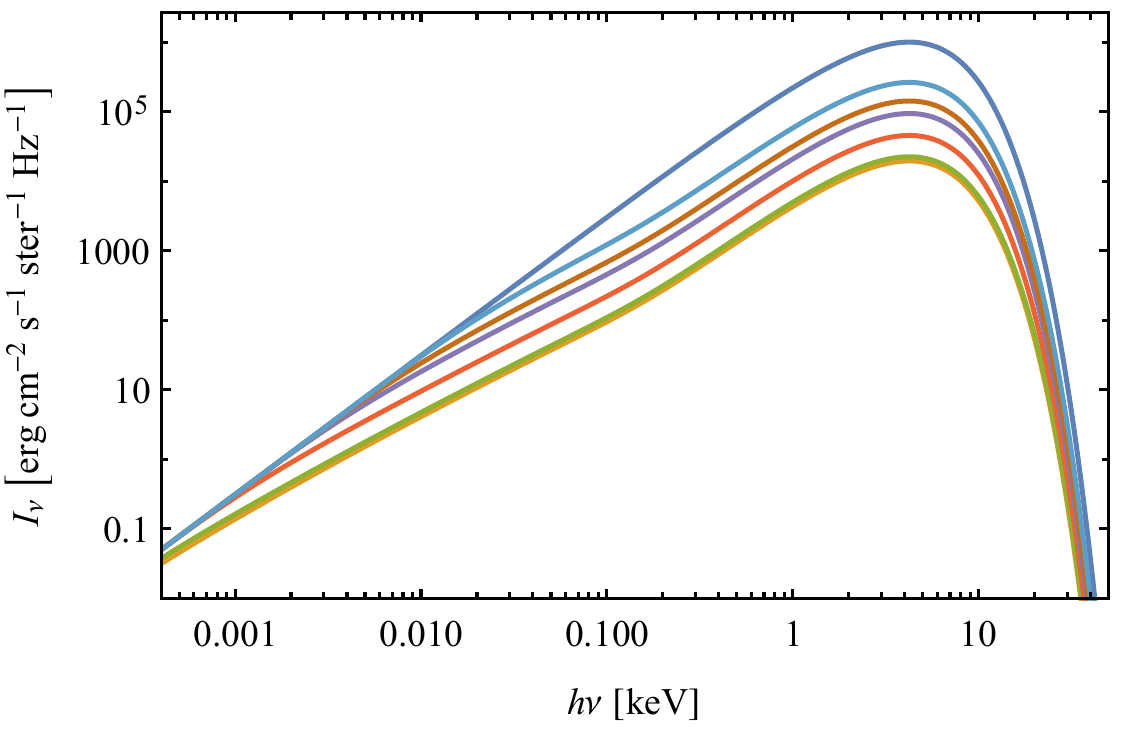}
   \\
        \includegraphics[width=0.23\textwidth]{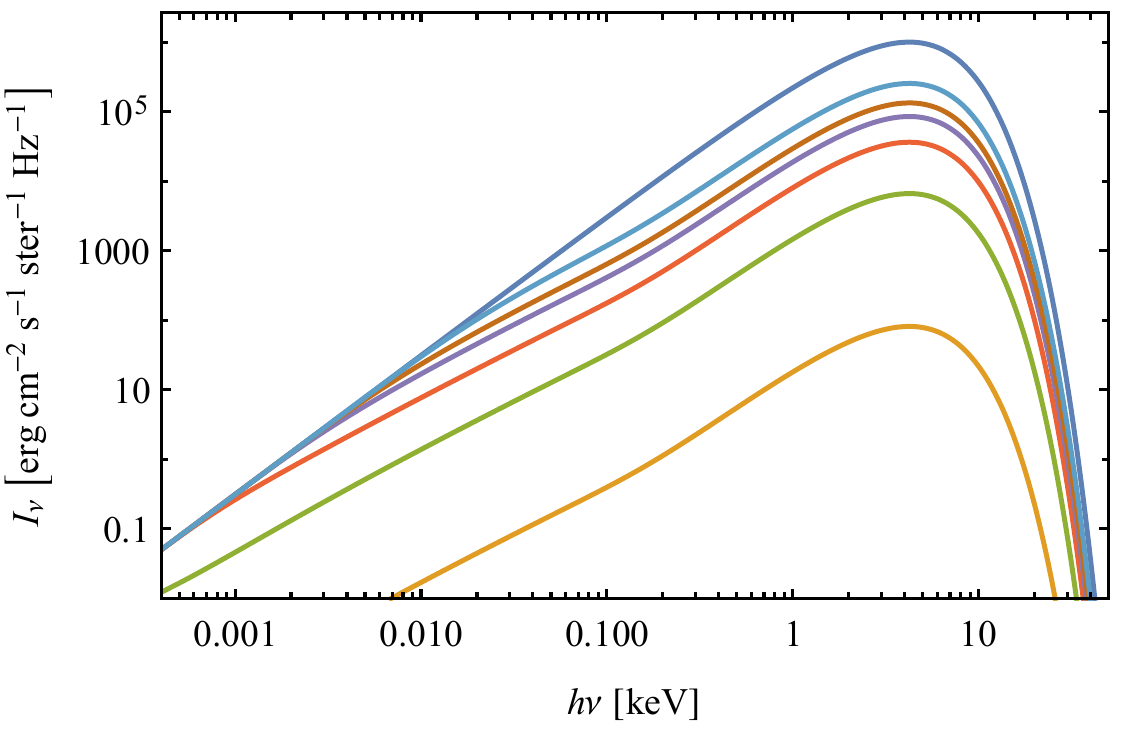}
  \hfill
                                 \includegraphics[width=0.23\textwidth]{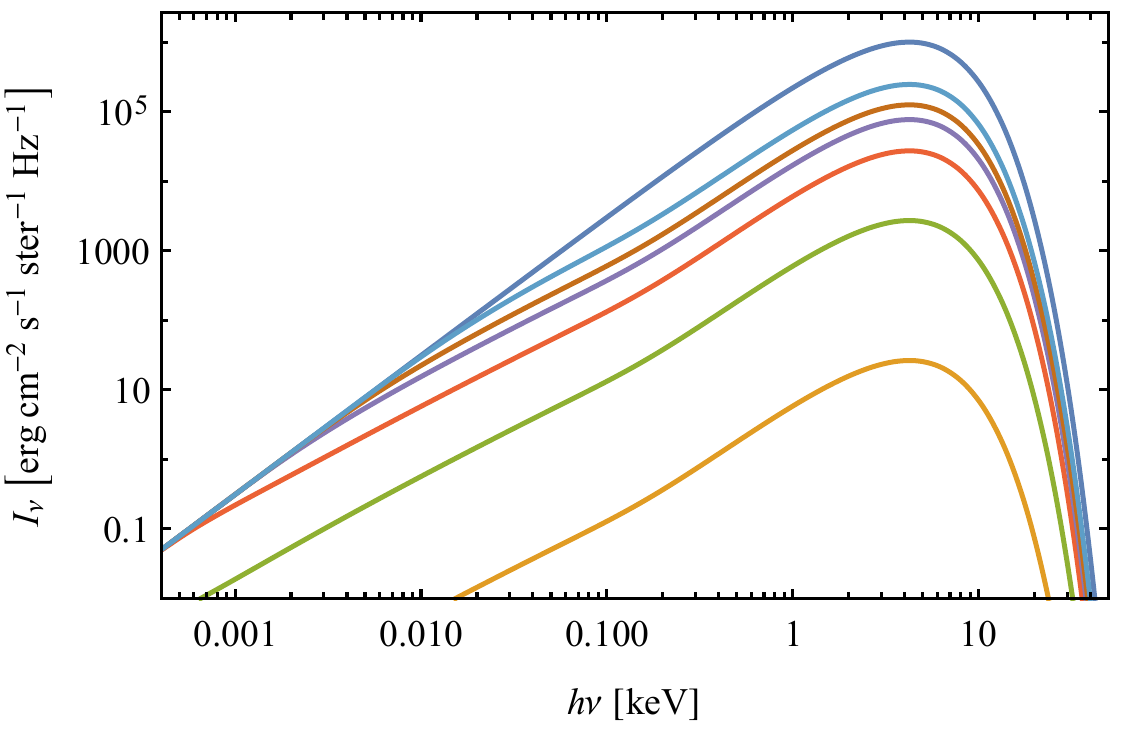}
  \hfill                               
                                   \includegraphics[width=0.23\textwidth]{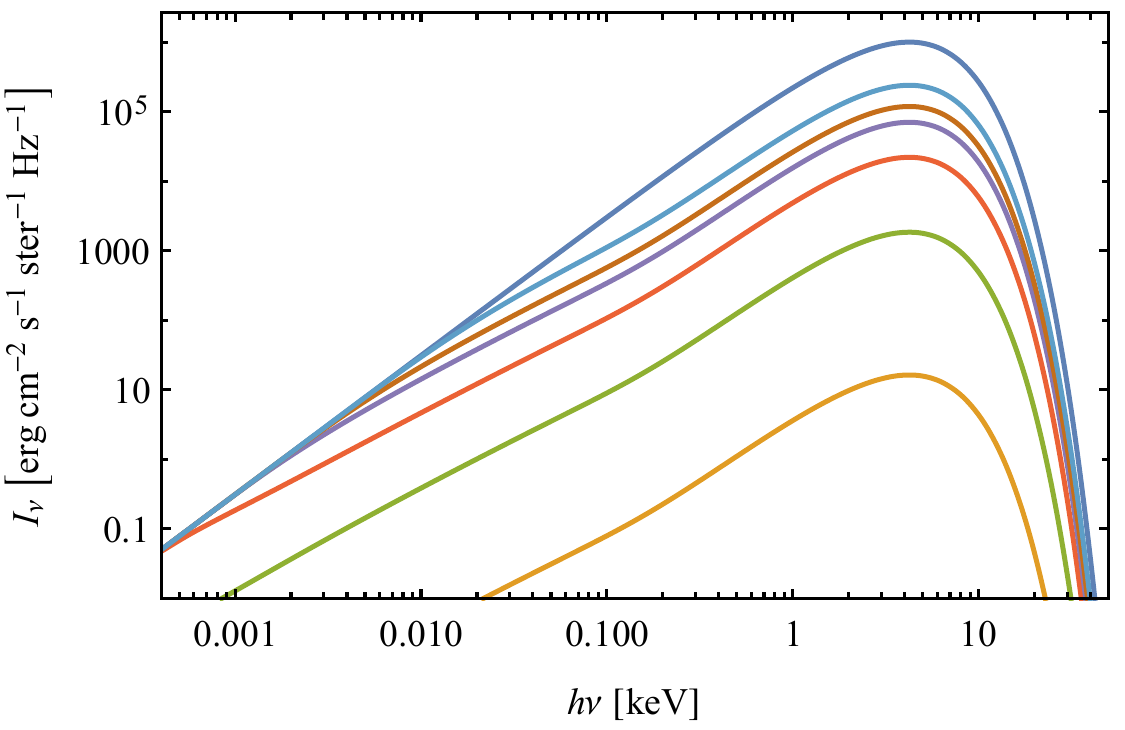}
  \hfill
                                 \includegraphics[width=0.23\textwidth]{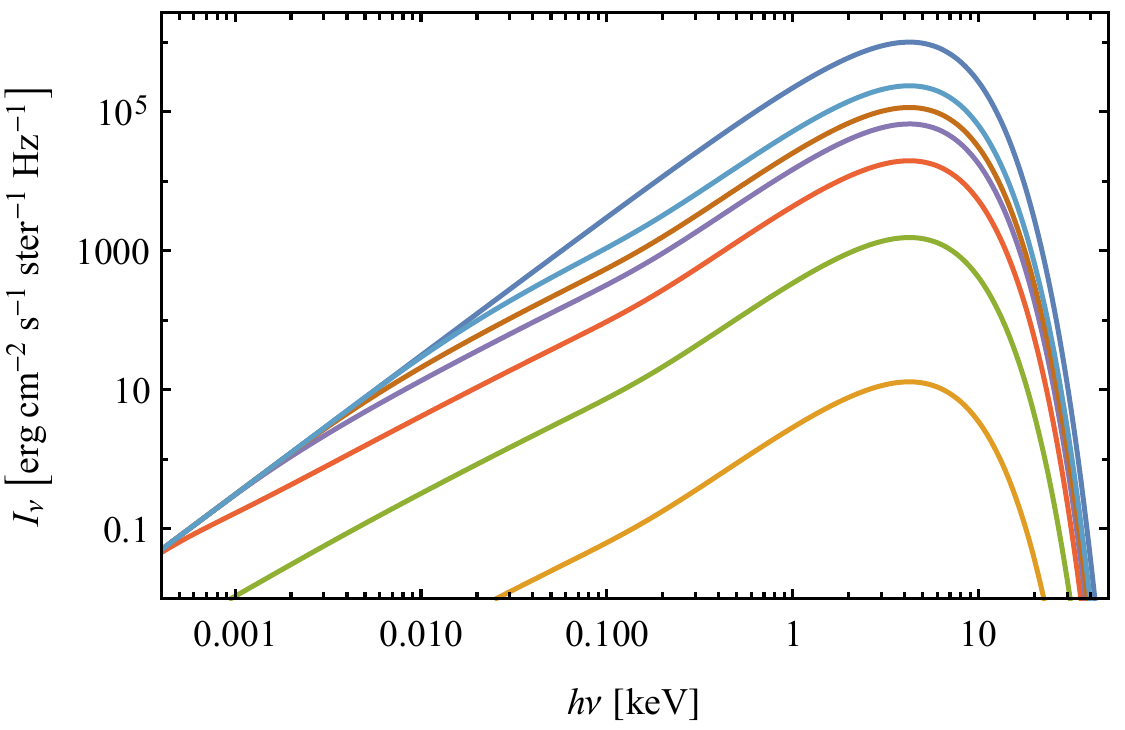}
    \\
   (b) 
  \\
  		\caption{The intensity $I_\nu(\mu)$ computed  due to numerical solving \eq{eq:transfm} for different $\tau$ and for different $\mu$ equal to (for the panels counted from left to right, from top to bottom, correspondingly)   $\mu_1$, $\mu_2$, $\mu_3$, $\mu_4$, 
  		$\mu_5$, $\mu_6$, $\mu_7$, $\mu_8$. The value of $\tau$: 0.001 (light brown), 0.1 (light green), 1 (red), 3 (purple), 5 (brown), 10 (blue), $\infty$ (dark blue).
  		The problem parameters:  \mbox{$T=1.5~{\rm keV}$}, \mbox{$M=1.5M_\odot$} and \mbox{$R=10^6~{\rm cm}$} (a), 
  		\mbox{$T=1.5~{\rm keV}$}, \mbox{$M=0.6M_\odot$} and \mbox{$R=7\cdot 10^8~{\rm cm}$} (b).
  		}
  \label{fig:Imu}
	\end{center}
 \end{figure*}

Discussing the transfer in terms of intensity it is useful to look at Fig.~\ref{fig:Imu}. The frequency dependencies of $I_\nu$ are computed at different $\tau$ for different values of cosine of the polar  angle    $\mu_1\approx 0.960$,   $\mu_2\approx 0.797$,  $\mu_3\approx 0.526 $, $\mu_4\approx 0.183 $,  $\mu_5\approx -0.183 $,   $\mu_6\approx -0.526 $,   $\mu_7\approx -0.797$,   $\mu_8\approx -0.960$.

Near the surface, where the radiation flux is close to the kinetic value, as it should be, $I_\nu(|\mu|)\gg I_\nu(-|\mu|)$ for sufficiently large  $|\mu|$. 
It can be seen that the radiation propagating in the directions corresponding to $\mu>0$ yields the contribution to the mean intensity, which causes bends in the spectral distributions of $J_\nu$ at relatively low frequencies (Fig.~\ref{fig:Jnu}, Fig.~\ref{fig:Jnudiff}). These bends are once again a good indication of the ranges in which the medium is optically thin. 

\begin{figure*}
 	\begin{center}
       \includegraphics[height=0.34\textwidth]{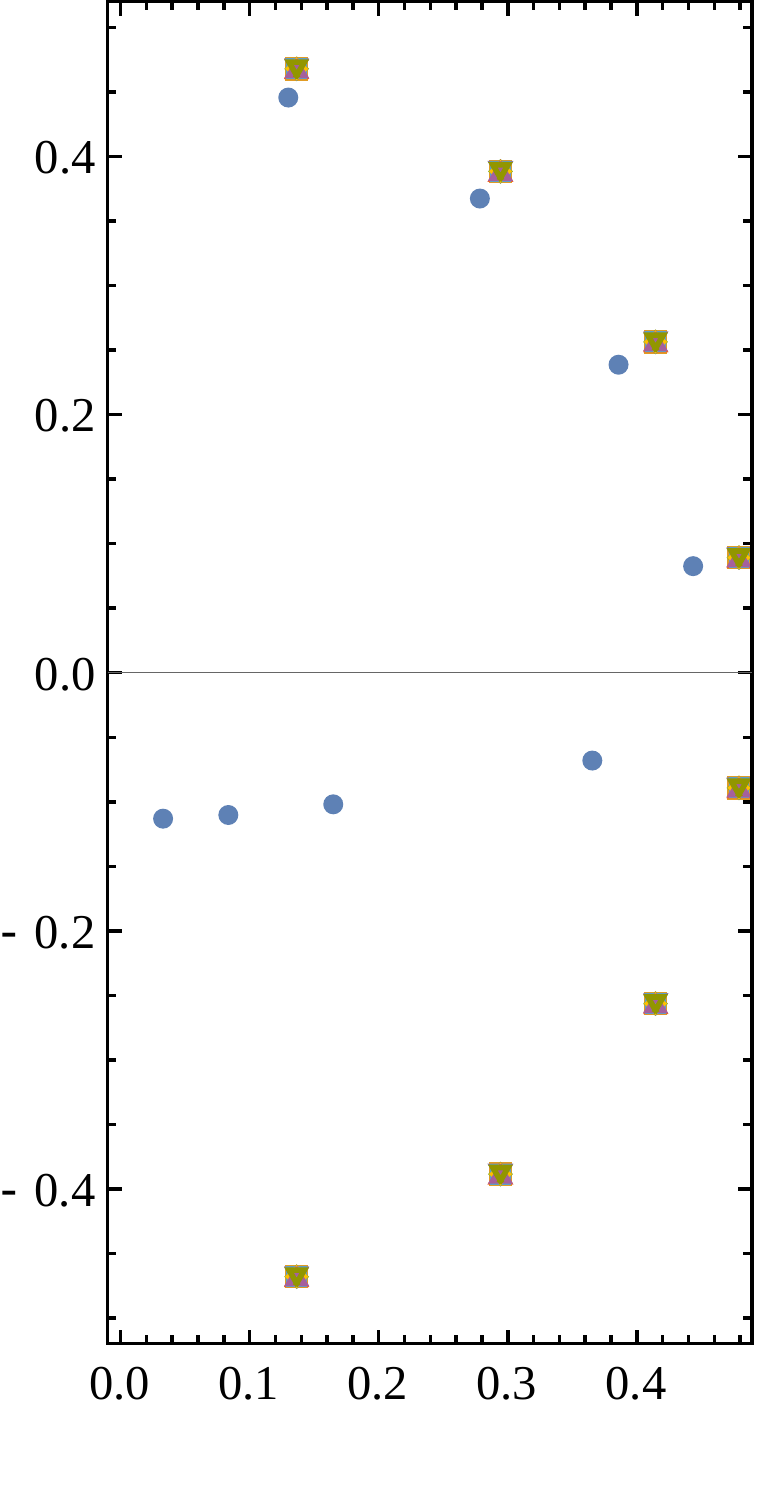}
  \hfill
   \includegraphics[height=0.34\textwidth]{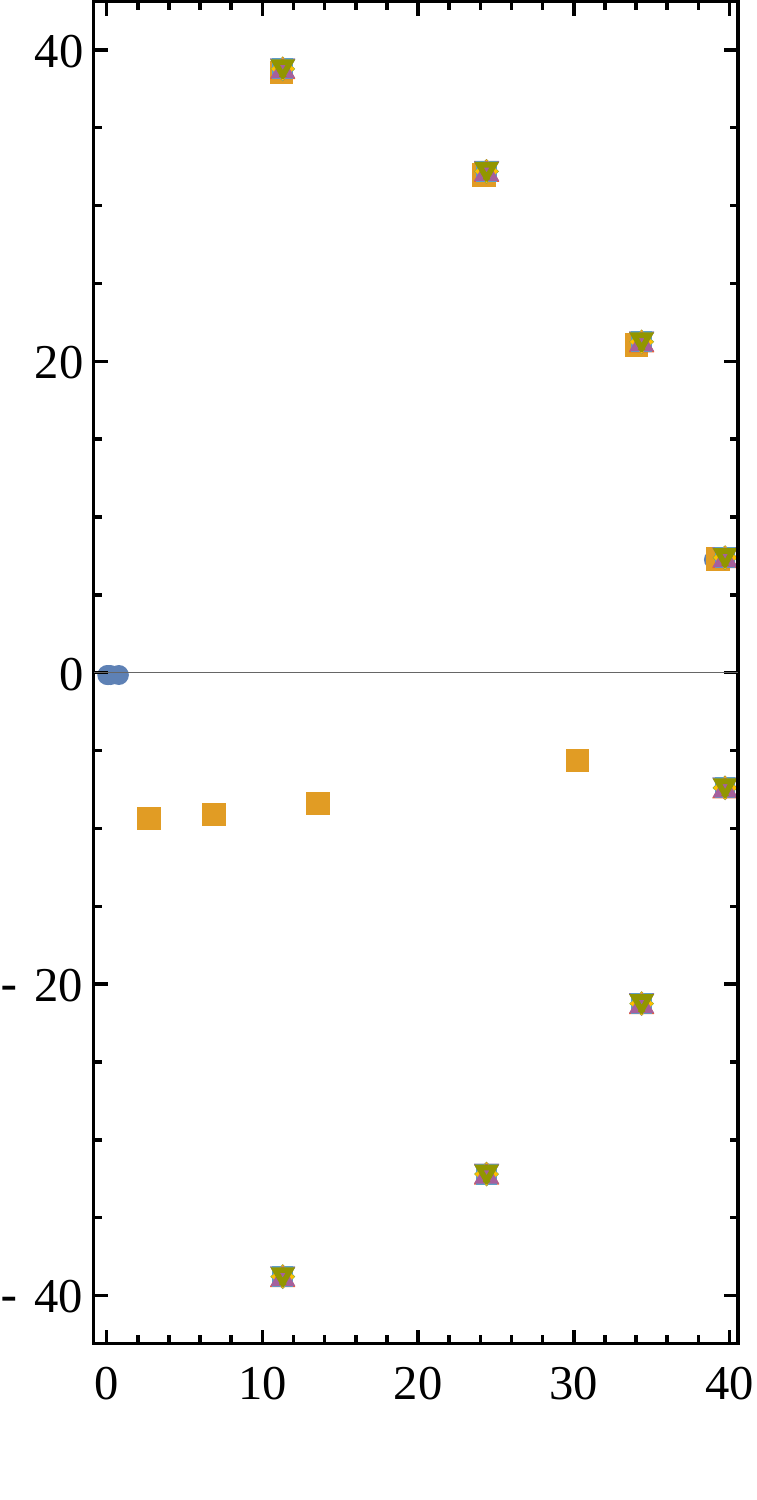}
   \hfill
     \includegraphics[height=0.34\textwidth]{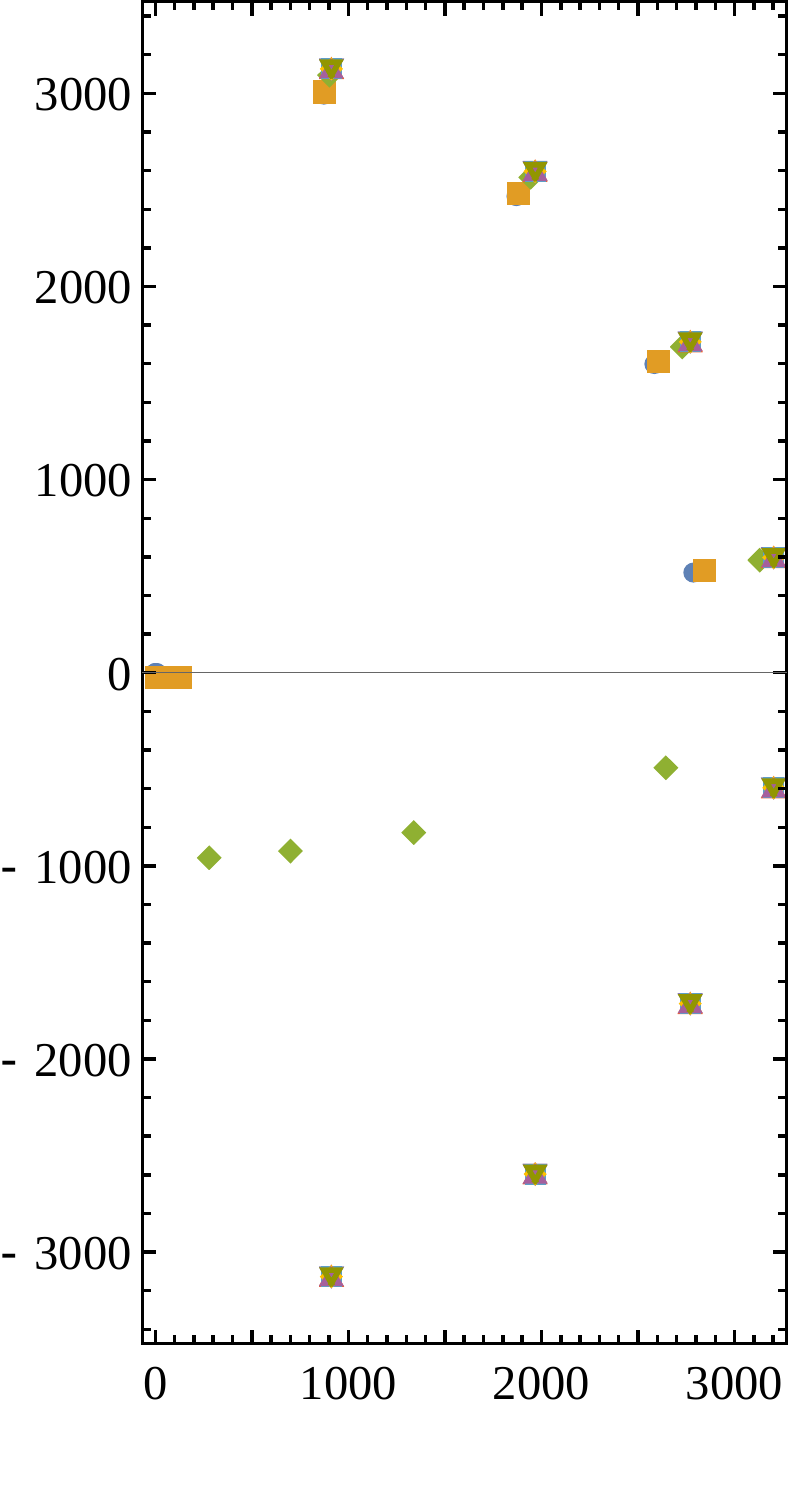}
  \hfill
  \includegraphics[height=0.34\textwidth]{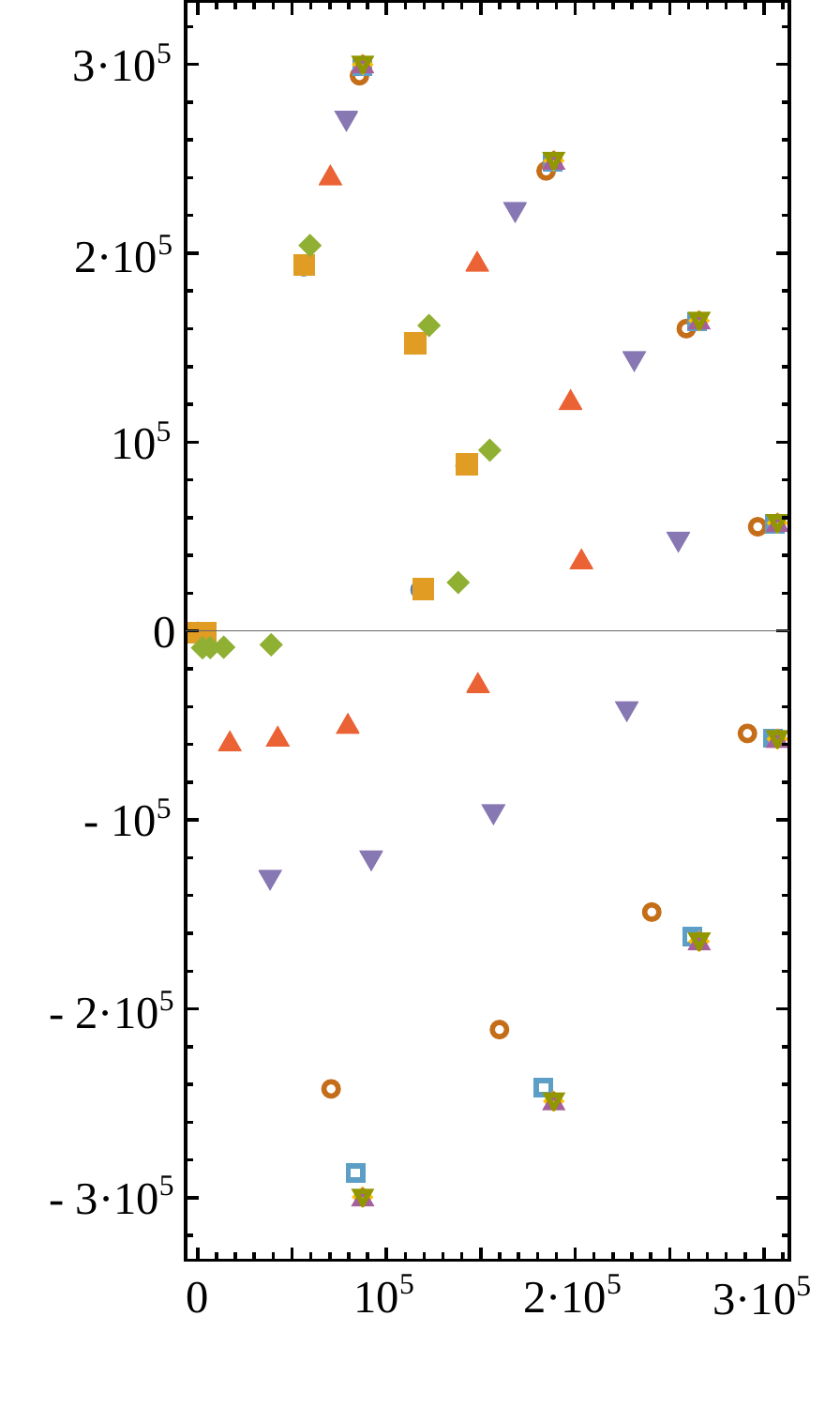}
    \hfill
    \includegraphics[height=0.34\textwidth]{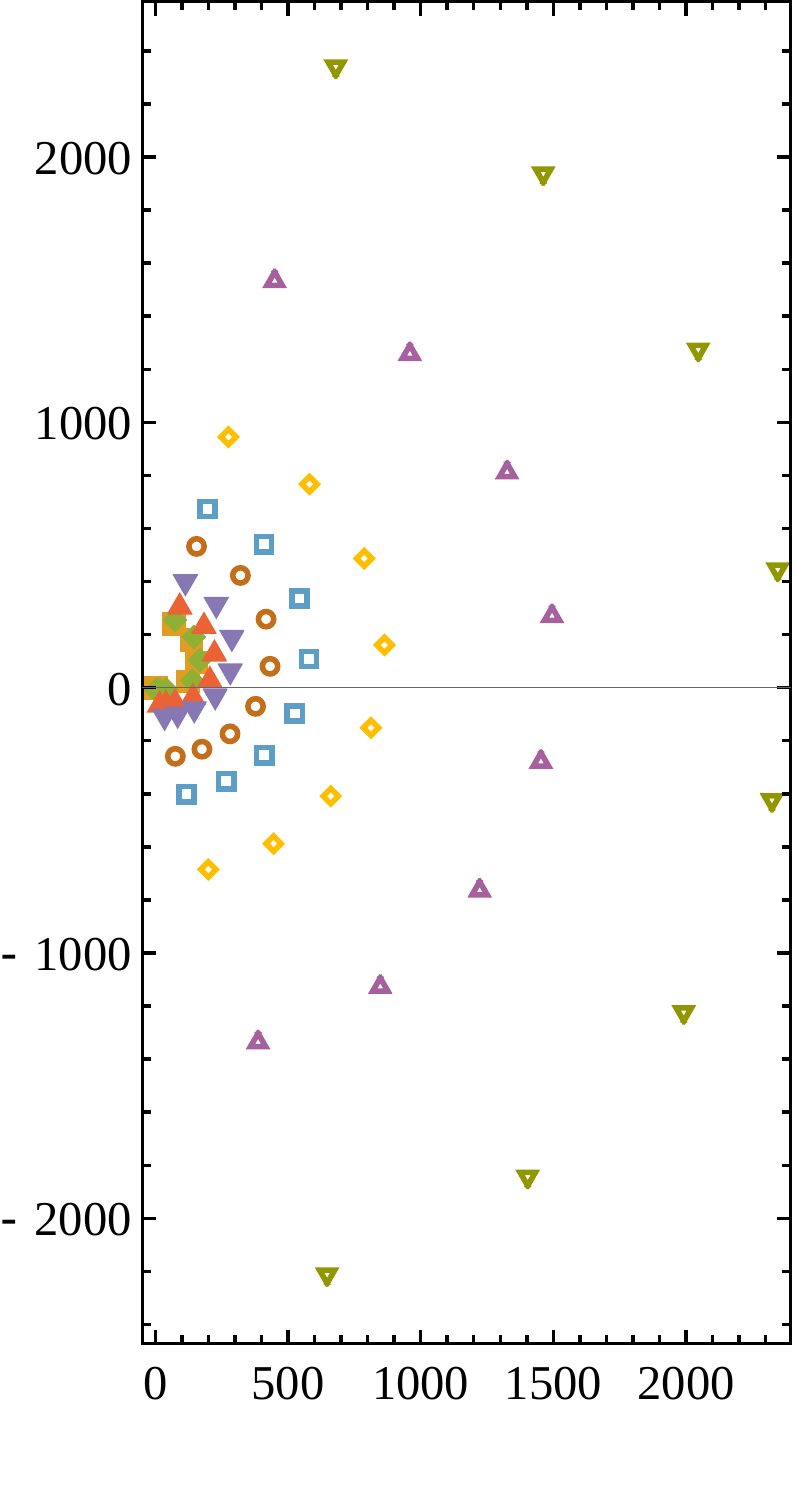}
    \\
   (a) 
  \\
   \vspace{15pt}
       \includegraphics[height=0.34\textwidth]{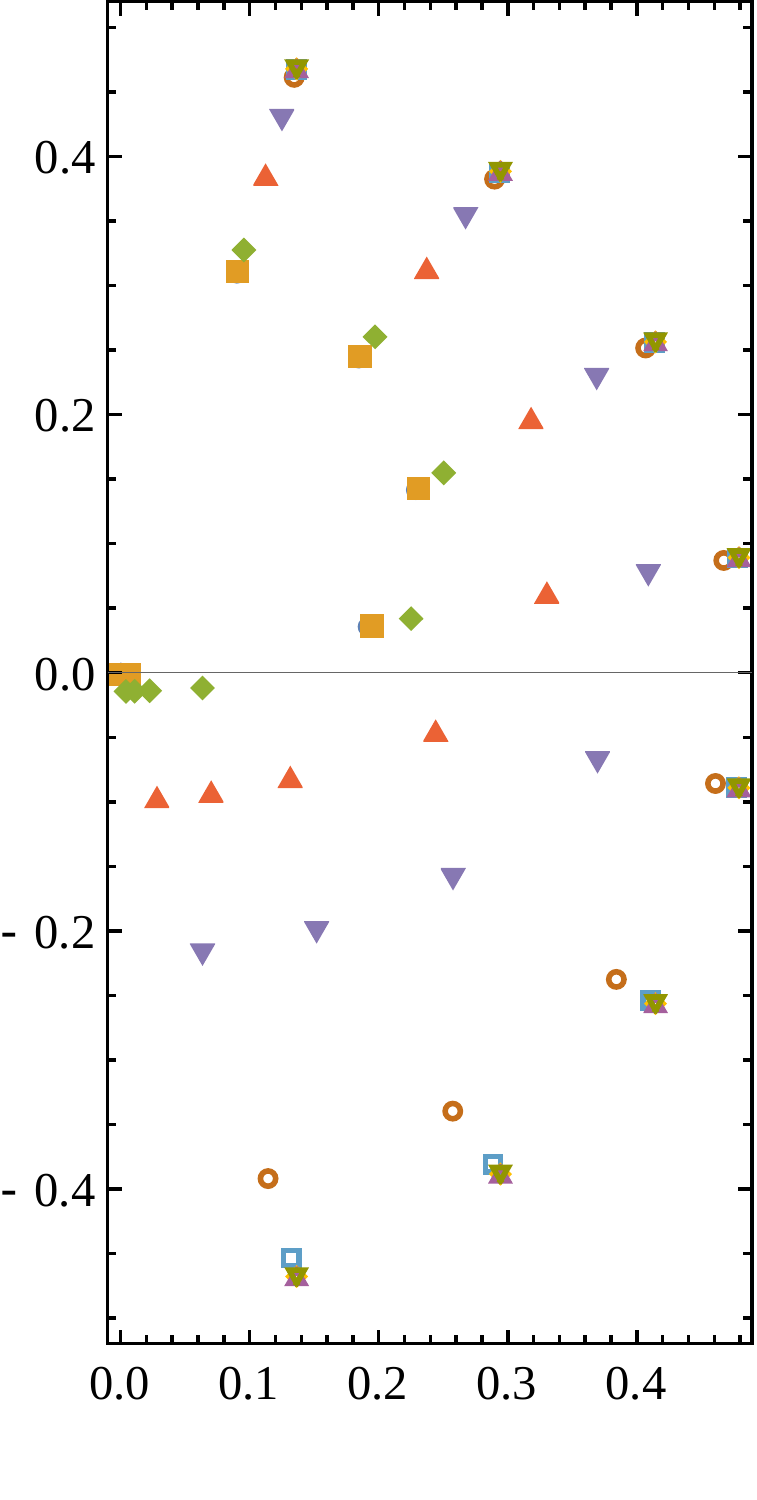}
  \hfill
  \includegraphics[height=0.34\textwidth]{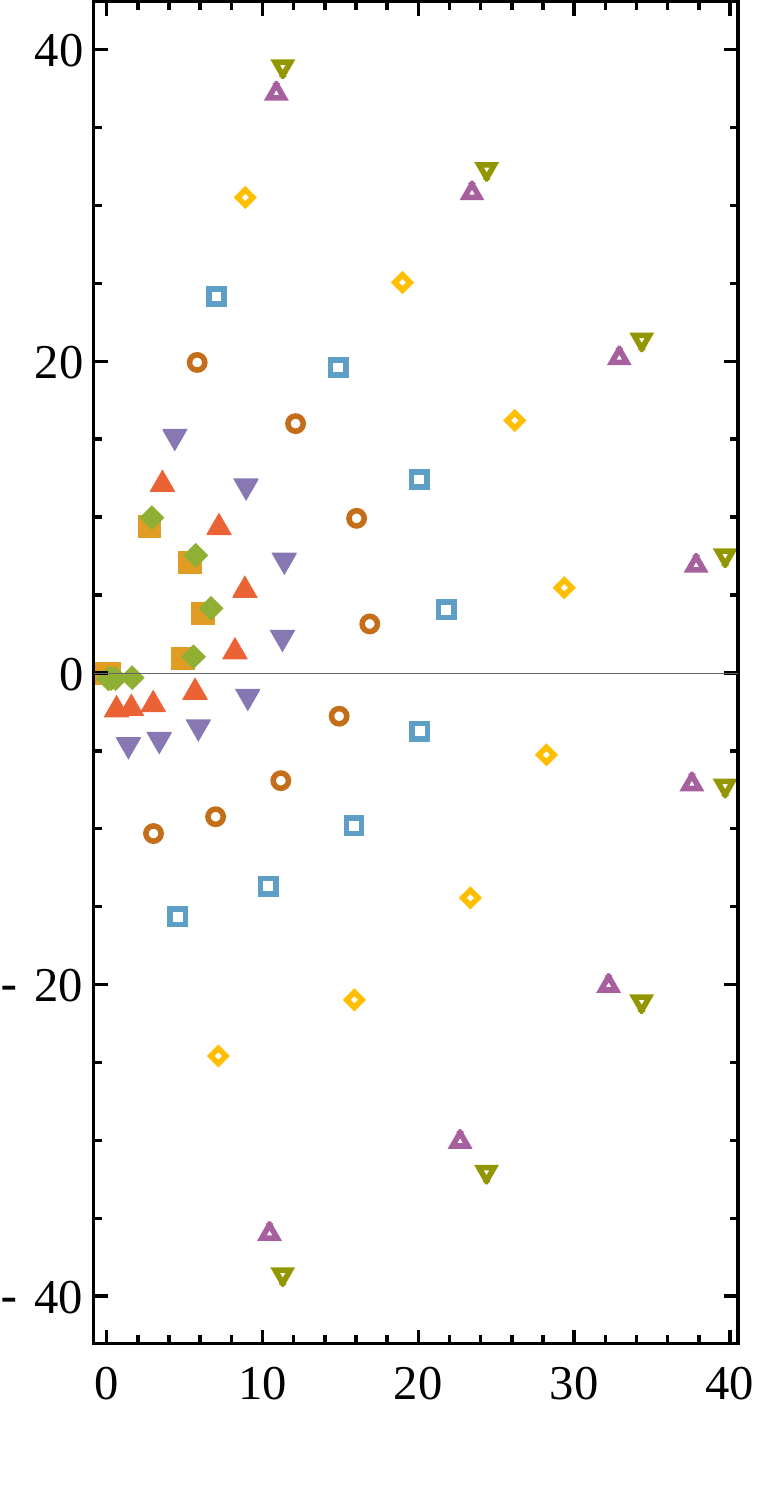}
                                  \hfill
     \includegraphics[height=0.34\textwidth]{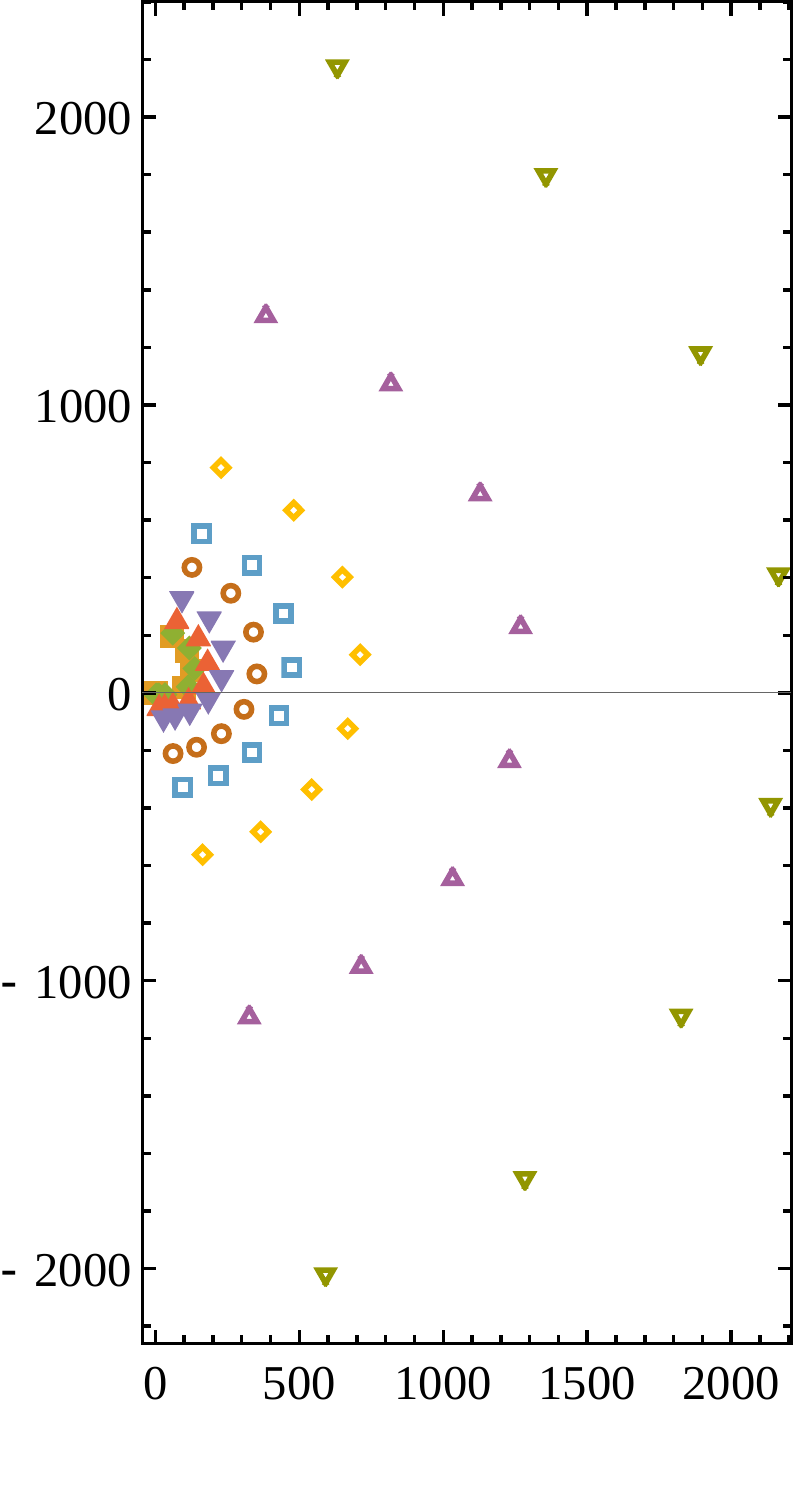}
  \hfill
\includegraphics[height=0.34\textwidth]{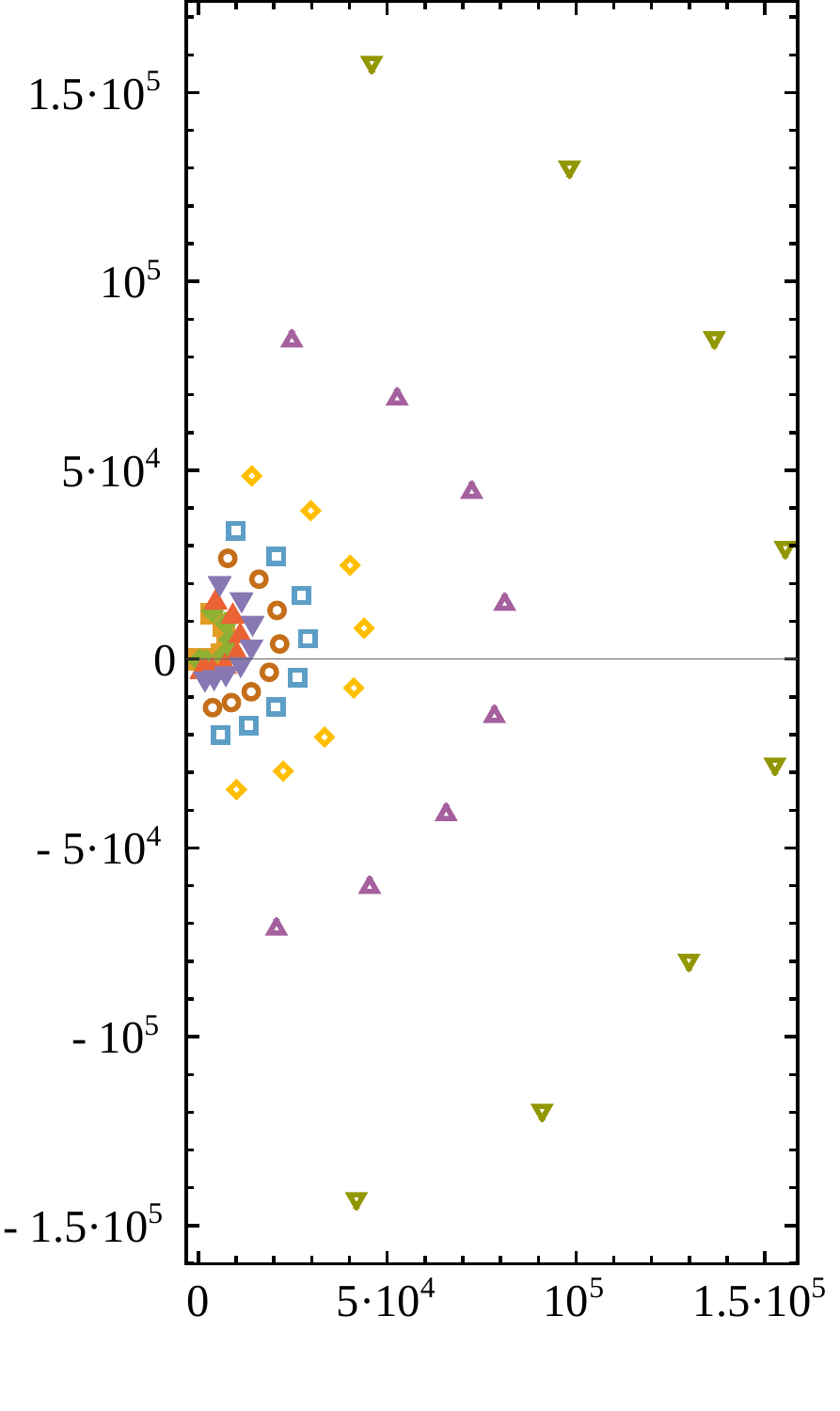}
        \hfill
\includegraphics[height=0.34\textwidth]{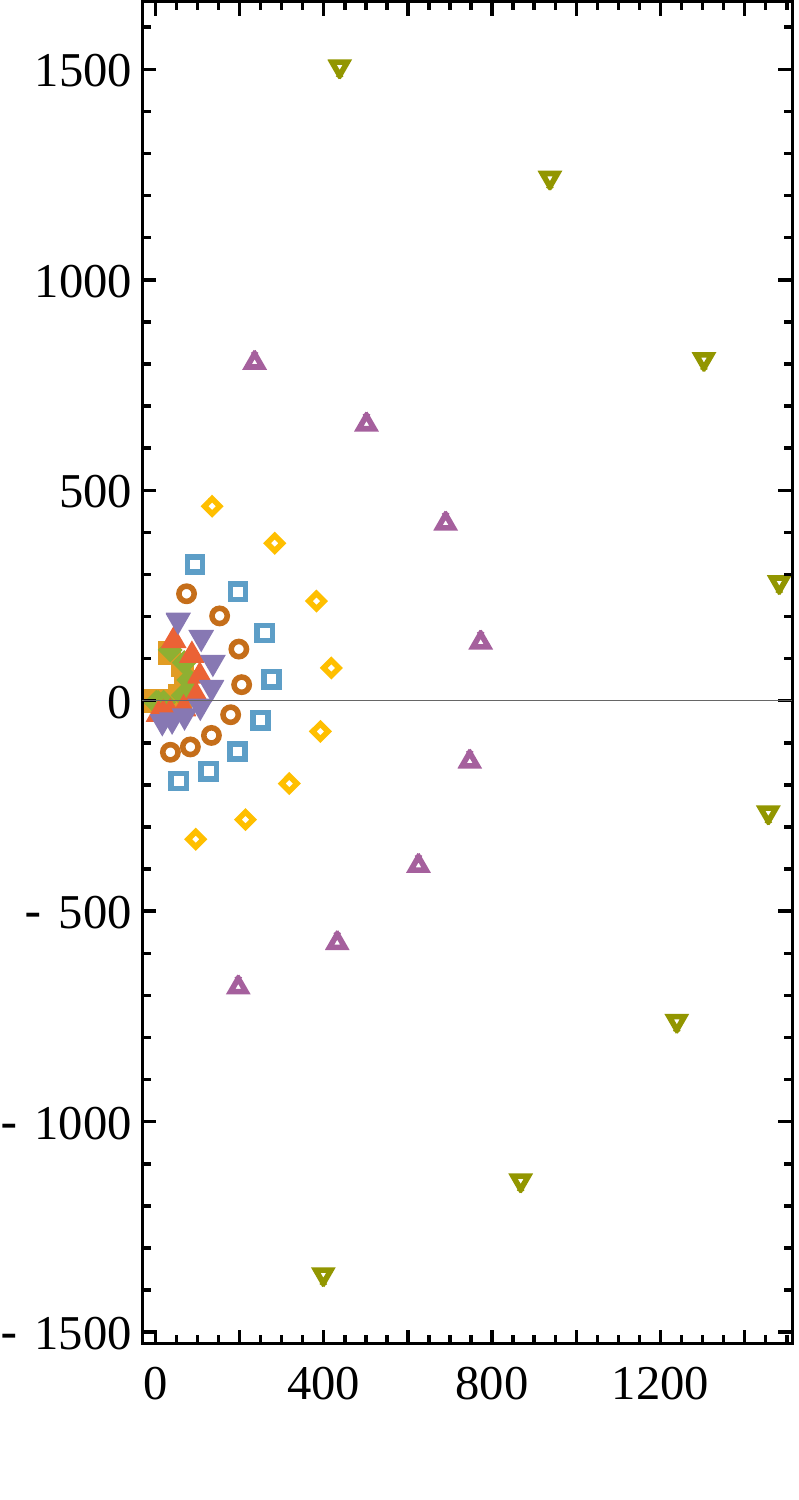}
                               \\
   (b) 
  \\   
   \vspace{5pt}     
  		\caption{Angular distributions of the intensity $I_\nu(\mu)$ (which is counted along the radius vector, 
  		erg~cm$^{-2}$~s$^{-1}$~ster$^{-1}$~Hz$^{-1}$) computed  due to numerical solving \eq{eq:transfm} for different $\tau$ and for different photon energies approximately equal to (the panels are counted from left) 0.001, 0.01, 0.1, 1.2, 20 keV. 
  		The values of  $\tau$ are: 0.001 (circles), 0.01 (squares), 0.1 (diamonds), 0.5 (red triangles),
  		1 (purple triangles), 2 (hollow circles), 3 (hollow squares), 5 (hollow diamonds), 10 (hollow purple triangles), 20 (hollow green triangles).
  		The problem parameters:  \mbox{$T=1.5~{\rm keV}$}, \mbox{$M=1.5M_\odot$} and \mbox{$R=10^6~{\rm cm}$} (a), 
  		\mbox{$T=1.5~{\rm keV}$}, \mbox{$M=0.6M_\odot$} and \mbox{$R=7\cdot 10^8~{\rm cm}$} (b). 
  		}
  \label{fig:Imup}
	\end{center}
 \end{figure*}

\begin{figure}
 	\begin{center}
  \includegraphics[width=0.48\textwidth]{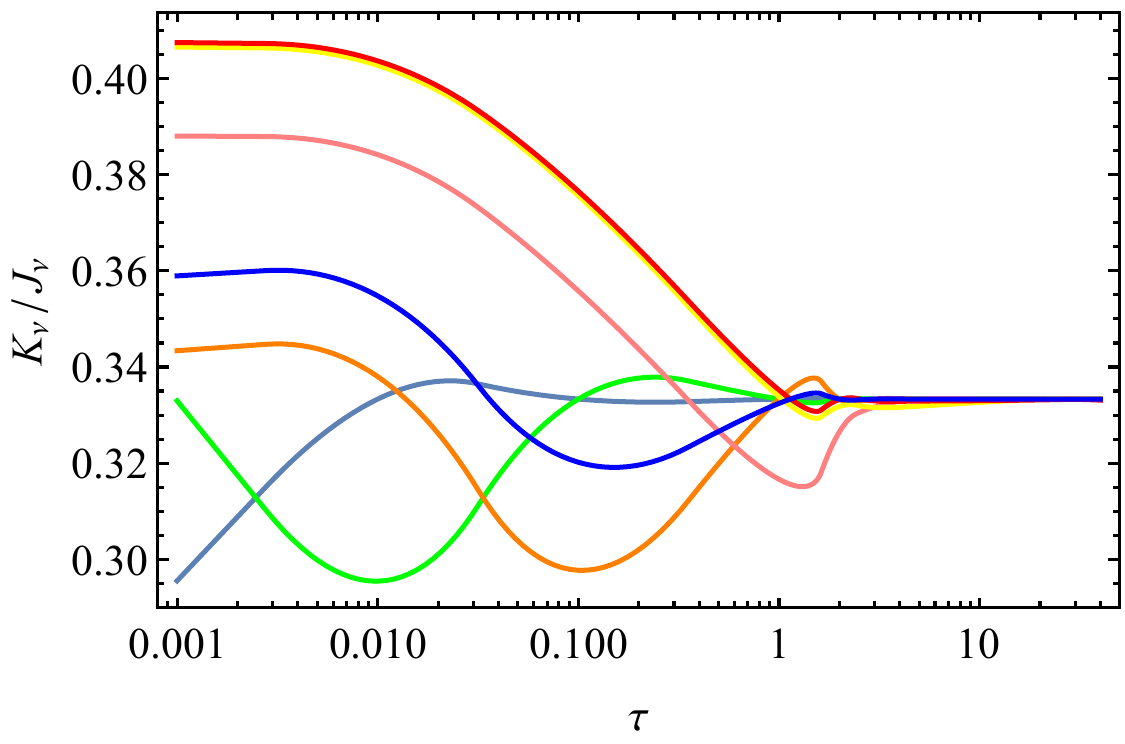}\\
 \vfill
  \vspace{1.2mm}
   \includegraphics[width=0.48\textwidth]{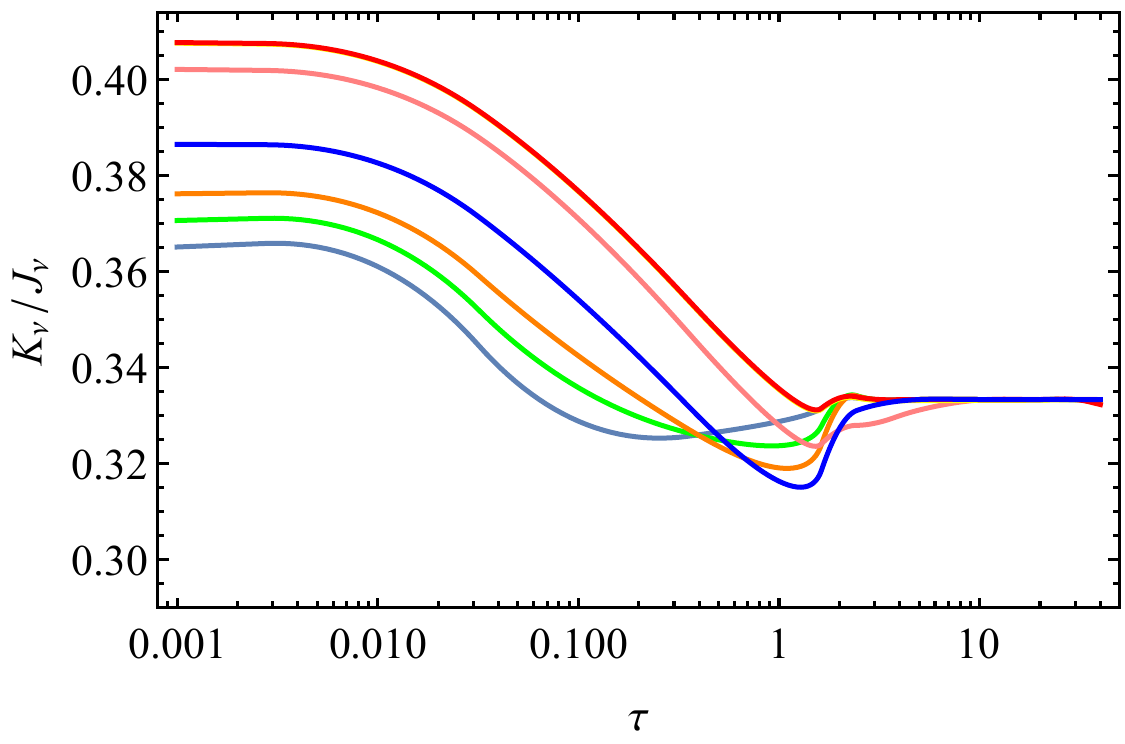}\\
  		\caption{The ratio $K_\nu/J_\nu$ as a function of $\tau$ plotted for different photon energies. These energies are approximately equal to \mbox{$1.3\cdot 10^{-3}$} (purple), $0.01$ (green), $0.1$ (orange), $0.3$ (blue),  $1.3$ (pink), 8.6 (yellow), 20 (red) keV for \mbox{$T=1.5~{\rm keV}$}, \mbox{$M=1.5M_\odot$} and \mbox{$R=10^6~{\rm cm}$} (upper panel), and $4\cdot 10^{-4}$ (purple), $5.5\cdot 10^{-4}$ (green), $7.2\cdot 10^{-4}$ (orange), $1.3\cdot 10^{-3}$ (blue),  $5\cdot 10^{-3}$ (pink), 0.1 (yellow), 20 (red) keV for \mbox{$T=1.5~{\rm keV}$}, \mbox{$M=0.6M_\odot$} and \mbox{$R=7\cdot 10^8~{\rm cm}$} (lower panel). }
  \label{fig:KJratio}%
	\end{center}
 \end{figure}

The angular distributions at various $\tau$ are shown in Fig.~\ref{fig:Imup} for several frequencies. It can be seen that an increase in the absorption cross-section towards low frequencies is expectedly accompanied by radiation isotropization at ever smaller $\tau$ (with equalization of the intensity values at different $\tau$ around blackbody value). In the case of a flatter density profile (the panels b), the values of the intensity tends to be compared outside the considered frequency range.
It is interesting to learn  the deviation of the ratio of the second intensity moment $K_\nu=\frac{1}{2}\int_{-1}^1\mu^2I_\nu {\rm d}\mu$ to the mean intensity from $1/3$. For the same multiple frequencies, the examples of results are shown in Fig.~\ref{fig:KJratio}. It can be seen that the distortion in the near-surface layers occurs differently depending on the value of the free-free absorption effects.




 \section{Remarks}
 \label{sec:remarks}
 
\subsection{Accreting atmosphere}

If  the atmosphere brakes the matter (for example, as in the case of X-ray pulsars at mass accretion rates low enough), 
the structure of the atmosphere is more complicated. 
The hydrostatic equilibrium is described then, generally speaking,   by  an equation that differs from \eq{eq:n_e}  and includes a term corresponding to the dynamic pressure. Moreover, in this situation the constant-temperature approximation becomes accurate only within the relatively cold layers lying under the matter stopping depth \linebreak \citep{1969AZh....46..225Z, 1982PlPh...24..339K}. The radiation escapes  from this zone not freely yet, and the escape is obstructed, apparently, by a 
layer of the Thomson optical thickness of several units, the temperature within which (especially in the upper part) is significantly higher than in the   region under consideration due to heating on account of the kinetic energy of the flow \citep{1969AZh....46..225Z}. In the hot layer one can not neglect the effect of Comptonization on the radiative transfer, although 
the Comptonization parameter (calculated with the Thomson opacity) of this layer should not exceed  unity in order of magnitude 
(calculated with taking into account the presence of a magnetic field, this parameter can only become  smaller, \citealt{1975A&A....42..311B}).
During the approximate considerations of  thermalization, it is possible therefore to use \eqn{eq:th2} omitting the presence of a hot layer.  
Accounting for the dynamic pressure, as well as using nonzero value of the thermal pressure as the boundary condition, 
leads to the cubic equation for~$\tau_{\rm th}$.

Simple estimations show that even in the case of a very strong magnetic field \mbox{$\sim 10^{12}$--$10^{13}$ G} the dynamic pressure can play a significant role only at sufficiently high accretion rates (under the condition that the Coulomb-braking regime takes place). The corresponding term depends  on the value of the matter flux
\beq{}
n_0 m_{\rm p}  V_{\rm ff}=\frac{\dot M}{ A},
\eeq
where $n_0$ is the electron number density at the boundary,  $V_{\rm ff}$  is the flow (free-fall) velocity, 
$\dot M$ is the  mass accretion rate at the considered emitting region, and $A$ is the total area of the accreting surface of this region ($n_0$ has thus an effective, averaged meaning because the flow is most likely inhomogenious).
Note that the mentioned term can depend in addition on the matter stopping depth and on the current value of the matter column density \citep{1969AZh....46..225Z}. 
The accretion on to white dwarfs being related with much lower surface magnetic field values can consequently be described  by the usual hydrostatic equilibrium equation.

\begin{figure}
 	\begin{center}
  \includegraphics[width=0.48\textwidth]{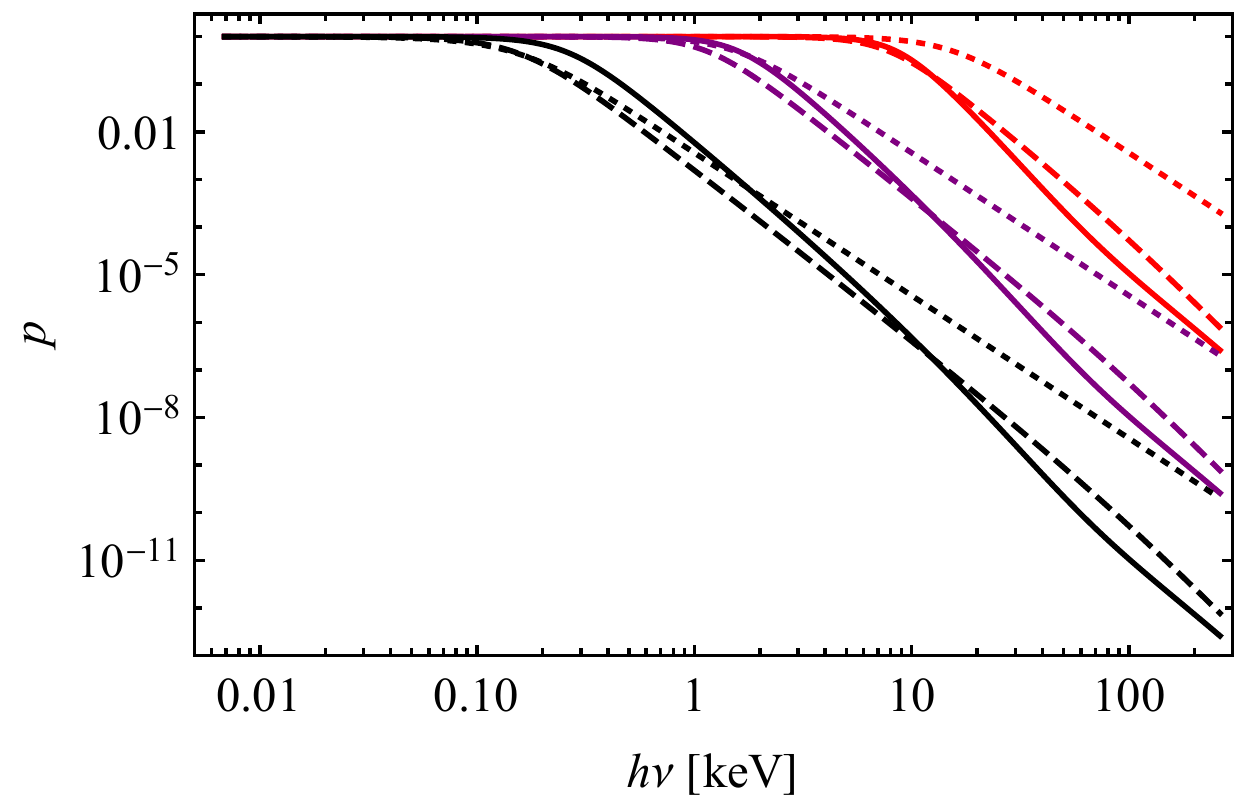}\\
  \vfill
  \vspace{1.2mm}
  \includegraphics[width=0.48\textwidth]{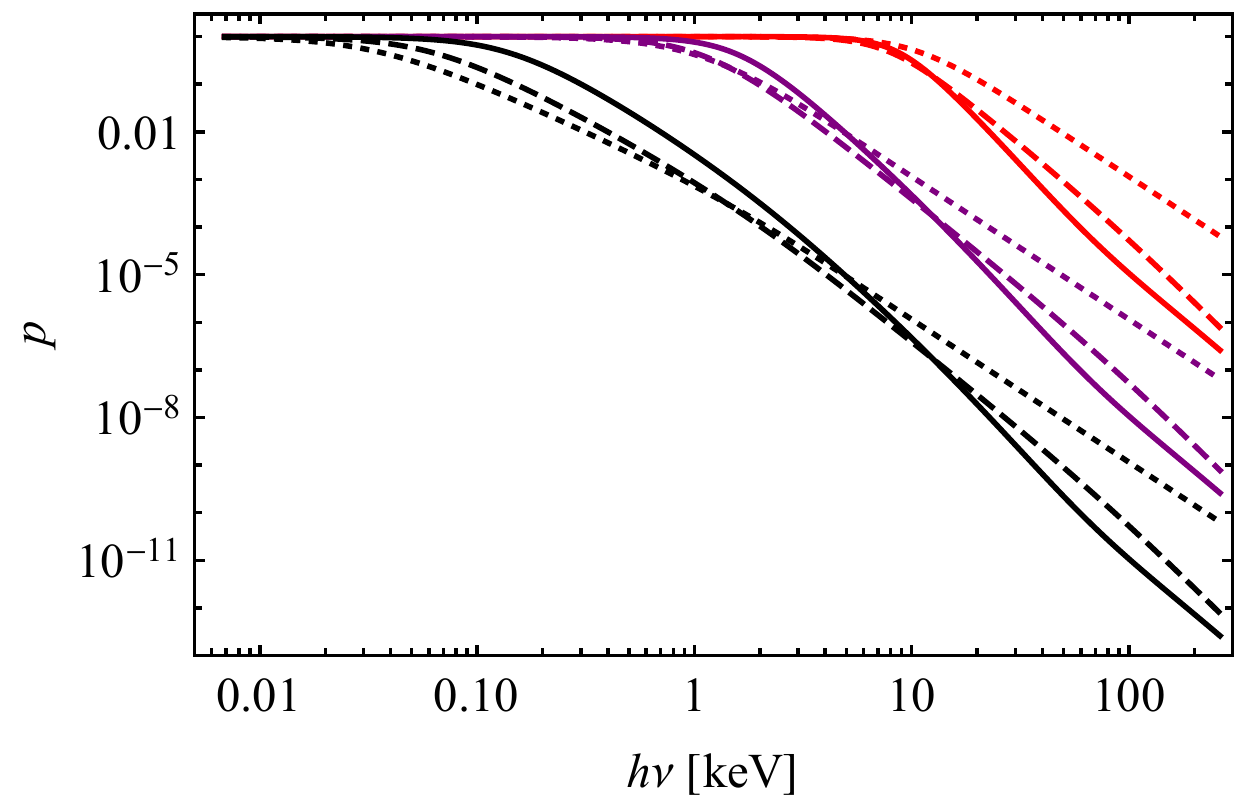}
  		\caption{
  		The probability of  free-free absorption in a cold magnetized plasma for the extraordinary mode (solid lines) and ordinary mode (dashed lines), Gaunt factors are not equal to unity. The dotted lines correspond to the computations of the absorption probability with the use of $\sigma_{\rm T}$ and nonmagnetized 
  		cross-section~\eqn{eq:abscorr}, $g_{\rm ff}=1$.     The electron  temperature: 
  		0.1~keV (upper panel),  1~keV (lower panel). The electron number density is equal to \mbox{$10^{20}~{\rm cm}^{-3}$} (black lines), 
  		\mbox{$10^{23}~{\rm cm}^{-3}$} (purple lines), \mbox{$10^{26}~{\rm cm}^{-3}$} (red lines). The magnetic field strength is  
  		\mbox{$5\cdot 10^{12}~{\rm G}$}.
  		}
  \label{fig:p}
	\end{center}
 \end{figure}

\subsection{Absorption probability in a cold magnetized plasma}

Now consider briefly the effects related with the presence of a strong magnetic field.  The behavior of the scattering and absorption cross-sections  in the dependence on the polarization of radiation, its frequency and on the direction of propagation of  photons with respect to the magnetic field was considered repeatedly.  
It is not difficult to obtain the absorption probability by means of the  magnetic (coherent) cross-sections calculated in the cold-plasma limit (e.g. \citealt{1979PhRvD..M}),
\beq{eq:pm}
p_i=\frac{\sigma_{{\rm a},\,i}}{\sigma_{{\rm s},\,i}+\sigma_{{\rm a},\,i}},~~i=1,~2,
\eeq
where $\sigma_{{\rm s},\,i}$ is the scattering cross-section in $i$th polarization mode (the sum of the cross-sections for scattering that occurs with and without the change of polarization), $\sigma_{{\rm a},\,i}$ is the free-free absorption cross-section in $i$th mode 
(see \citealt{1979ApJ...233L.125V}, where the expressions for the used cross-sections for said processes can be found).
The magnetic absorption cross-sections are normalized as specified in \cite{1992hrfm.book.....M} (see also \citealt{1961ApJS....6..167K}), and the Gaunt factors are computed in accordance with  \cite{nagel1980}.
Fig.~\ref{fig:p} shows the frequency dependencies for the absorption probability  for different values of the plasma density at fixed magnetic field strength and electron temperature (which is calculated in correspondence with the one-dimensional Maxwellian distribution for the case of magnetized plasma). 
The cross-sections are calculated in the weak dispersion limit with making use of the parameter   \mbox{$b=\nu_{\rm c}\sin^2\theta/(2 \nu \cos\theta)$}, where $\theta$ is the angle made by the wavevector with the magnetic field direction and $\nu_{\rm c}$  is the cyclotron 
frequency (thus, the parameter \mbox{$\delta=0$} and the parameter $v\ll 1$, which is also set to be equal to zero in computations, notations are usual). 

The effect of the redistribution for  polarized photons in a hot magnetized plasma was investigated, for example, by~\cite{nagel1981}.

\section{Conclusions}\label{sec:concl}
Investigating the problem of the radiative transfer under the conditions which are set in Section~\ref{sec:depth} shows that
it appears to be possible to describe the modification of the spectrum with the depth in terms of the characteristic frequency-dependent (optical) depth indicating the extent of thermalization of radiation (at a given point of space). Such a description  is not related with the certain shape of the spectrum  and thus has only a qualitative meaning from the clear point of view. In the case of a hydrostatic  atmosphere with the predominance of scattering (at considered frequencies),  \eq{eq:th2} leads to the  linear dependence of such a characteristic optical depth on the frequency of  radiation. 
This depth is determined quantitatively by the expression~\eqn{eq:tau_th}. Comparison with the values found from the solutions of the transfer equation suggests that the accuracy of this estimate is about a factor  \mbox{$\approx 2$} (the Gaunt factor has not yet been taken into account either in the probabilistic definition or in the transfer equations~-- that should not be dramatic, see, e.g., fig.~3 in \citealt{1961ApJS....6..167K}).
The dependencies of the depth $\tau^*_{\rm th}$, which is found from these solutions, on the temperature in the atmosphere, and on the mass and radius of a gravitating object  are in accordance with~\eq{eq:tau_th} in the domain of the parameters where comparing is appropriate.

It is reasonable to discuss the radiation thermalized `partially'  only within the appropriate ranges of the problem parameters, obviously. That part of the emergent radiation spectrum which corresponds to the main contribution to  the emergent  radiation energy density must lie in the frequency range where scattering prevails with respect to absorption in the region of the spectrum formation~-- the condition that the \mbox{X-ray} radiation satisfies for the typical parameters of the neutron star, for example. It is clear that in the case of the atmosphere of a neutron star,  temperatures of about 0.1~keV already  allow  radiation to become thermalized practically at once (around some $\tau$) at all frequencies with increasing depth. In the case of  white dwarfs, the linear dependence of $\tau_{\rm th}$ on frequency is explicitly held over a wider temperature range.

One might point to the definition of the thermalization depth used in astrophysics, which is related to the expression $1/\sqrt{p}$. Such a definition comes from the form of analytical solutions of  \eq{eq:transf2}  including the photon destruction probability $p$ expression for which is not specified (e.g.~\citealt{1982stat.book.....M}) and often arise due to describing the transfer in spectral lines formed in stellar atmospheres (e.g.~\citealt{sobolev}). This expression for the thermalization depth, as a rule, is also considered to satisfy the statement about the thermalization of a photon for $1/p$ scatterings. But this is the expression, which  above was taken as the basis for the very definition of the thermalization depth (mainly for the absorption coefficient of a specific form, for which one can write the separate solutions of the transfer equation). Although in our case thus there is no need to use the expression $1/\sqrt{p}$, one may check the degree of consistency of the solutions. Denoting $\tau_{\nu,\, \rm th}(\tau)=1/\sqrt{p}$, it is possible to write for the case of $p\propto \tau$ that
\beq{}
\frac{C_1}{2\sigma_{\rm T}}\tilde\tau_{\rm th}^2+\tilde\tau_{\rm th}=\sqrt{\frac{\sigma_{ \rm T}}{C_1\tilde\tau_{\rm th}}},
\eeq
where $\tilde\tau_{\rm th}$ is the solution for the  Thomson optical depth corresponding $\tau_{\nu, \,\rm th}$. To analyze the solution, it is easier to obtain the dependence of the characteristic frequency on $\tau$. Having the cumbersomeness form, such a solution coincides with found from \eqn{eq:tau_th} at sufficiently high energies ($>2~{\rm keV}$ for typical neutron star parameters) and it loses its accuracy in comparison with $\tau_{\rm th}$ and $\tau_{\rm th}^*$ at low frequencies, 
where the dependence  $\tilde\tau_{\rm th}(\nu)$ goes considerably below the corresponding linear one.

The behaviour of quantity $\tau_{\rm th}$ (and $\tau^*_{\rm th}$) can be considered as the  property of  solutions \eqn{eq:J} and \eqn{eq:Jabs} of the radiative transfer equation of the form \eqn{eq:transf} and \eqn{eq:transf2}, it seems to be useful in theoretical deliberations. 
The obtained numerical results can be further refined by considering more precise expressions for the free-free absorption cross-section (taking into account the distinction of the Gaunt factor from unity), and also by considering more accurate expressions for the source function,  which would not imply its isotropy.
Namely, one can take into account the redistribution over angles and frequencies during scattering and, furthermore, the distinction of the emission coefficient term depending on absorption cross-section   from the value used above and determined by the Planck spectrum (including  accounting the angular dependence of this term).
  
The circumstances considered in this communication   are notably idealized, but applications of the results, at least for estimates, 
can be found: one of the examples is the atmospheres of accreting highly magnetized neutron stars.

 \bibliographystyle{mnras}
\bibliography{thd}

\label{lastpage}
\end{document}